\documentclass[a4paper,fleqn,12pt]{cas-sc}

\usepackage{geometry}
\usepackage[numbers,sort]{natbib}

\usepackage{afterpage}
\usepackage{soul}
\usepackage{underscore}
\usepackage{esint}
\graphicspath{ {figs/} {tabs/} {thumbnails/} }
\usepackage{subfig}
\usepackage[dvipsnames]{xcolor}
\usepackage{bm}
\usepackage{lipsum} 
\usepackage{cleveref}
\usepackage{fancyhdr}
\crefname{figure}{Fig.}{Figs.}%
\crefname{equation}{Eq.}{Eqs.}%
\crefname{table}{Tab.}{Tabs.}%
\usepackage{expl3}
\usepackage{enumitem}
\usepackage{calc} 
\usepackage{amsthm,amsmath}
\usepackage[flushleft]{threeparttable}
\usepackage{pdfpages}
\usepackage[dvipsnames]{xcolor}
\usepackage{pdflscape}

\usepackage{algorithm}
\usepackage{algpseudocode}

\algdef{SE}[REQUIRE]{Require}{EndRequire}{\textbf{Entrée:}}{\textbf{Fin}}
\algdef{SE}[ENSURE]{Ensure}{EndEnsure}{\textbf{Sortie:}}{\textbf{Fin}}

\ExplSyntaxOn

\cs_new_protected:Npn \__create_math_cmd:Nnn #1#2#3 {
  \cs_new:cpn {#2#3} {#1{#2}}
}

\clist_map_inline:nn { A,B,C,D,E,F,G,H,I,J,K,L,M,N,O,P,Q,R,S,T,U,V,W,X,Y,Z }
  {
    \__create_math_cmd:Nnn \mathcal {#1} {cal}
  }

\clist_map_inline:nn { A,C,D,E,F,G,H,I,J,K,L,M,N,O,P,Q,R,S,T,U,V,W,X,Y,Z }
  {
    \__create_math_cmd:Nnn \mathbb {#1} {bb}
  }

\clist_map_inline:nn { A,C,D,E,F,G,H,I,J,K,L,M,N,O,P,Q,R,S,T,U,V,W,X,Y,Z }
  {
    \__create_math_cmd:Nnn \mathfrak {#1} {frak}
  }

\clist_map_inline:nn { A,B,C,D,E,F,G,H,I,J,K,L,M,N,O,P,Q,R,S,T,U,V,W,X,Y,Z }
  {
    \__create_math_cmd:Nnn \boldsymbol {#1} {bs}
  }

\clist_map_inline:nn { a,b,c,d,e,f,g,h,i,j,k,l,m,n,o,p,q,r,s,t,u,v,w,x,y,z }
  {
    \__create_math_cmd:Nnn \boldsymbol {#1} {bs}
  }

\clist_map_inline:nn { A,B,C,D,E,F,G,H,I,J,K,L,M,N,O,P,Q,R,S,T,U,V,W,X,Y,Z }
  {
    \__create_math_cmd:Nnn \mathbf {#1} {bf}
  }

\clist_map_inline:nn { a,b,c,d,e,f,g,h,i,j,k,l,m,n,o,p,q,r,s,t,u,v,w,x,y,z }
  {
    \__create_math_cmd:Nnn \mathbf {#1} {bf}
  }

\ExplSyntaxOff

\newcommand{\onebf}{\mathbf{1}}
\newcommand{\Psibf}{\boldsymbol{\Psi}}
\newcommand{\alphabf}{\boldsymbol{\alpha}}

\newcommand{\Sigmabf}{\boldsymbol{\Sigma}}
\newcommand{\Phibf}{\boldsymbol{\Phi}}
\newcommand{\transpose}{^\intercal}
\newcommand{\matlab}{M\scriptsize ATLAB \normalsize}
\newcommand{\uqlab}{UQL\scriptsize ab \normalsize}

\renewcommand{\leq}{\leqslant}

\renewcommand{\geq}{\geqslant}
\renewcommand{\transpose}{^\top}
\newtheorem*{remark}{Remark}
\newtheorem*{prop}{Proposition}

\usepackage{nomencl}
\usepackage{ifthen}
\makenomenclature

\renewcommand{\nomgroup}[1]{%
  \item[\bfseries%
  \ifthenelse{\equal{#1}{A}}{Abbreviation}{%
  \ifthenelse{\equal{#1}{B}}{Latin capital}{%
  \ifthenelse{\equal{#1}{C}}{Latin lowercase}{%
  \ifthenelse{\equal{#1}{D}}{Greek capital}{%
  \ifthenelse{\equal{#1}{E}}{Greek lowercase}{%
  \ifthenelse{\equal{#1}{F}}{Calligraphic}{%
  \ifthenelse{\equal{#1}{G}}{Blackboard bold}{%
  }}}}}}}]%
}
\usepackage{letltxmacro}
\usepackage{graphicx}

\LetLtxMacro{\oldtable}{\table}
\LetLtxMacro{\endoldtable}{\endtable}

\renewenvironment{table}[1][htbp]{%
  \oldtable[#1]%
  \centering
  \small
  \begin{lrbox}{0}%
    \begin{minipage}{1.1\textwidth}%
}{%
    \end{minipage}%
  \end{lrbox}%
  \resizebox{\textwidth}{!}{\usebox{0}}%
  \endoldtable
}

\makeatletter
\AtBeginDocument{
\def\ps@first{%
  \let\@oddhead\@empty
  \let\@evenhead\@empty
  \def\@oddfoot{\small\sffamily J. Minini \& M. Wasem \hfill Page~\thepage\space of~\lastpage}%
  \let\@evenfoot\@oddfoot
}
\def\ps@cas{%
  \let\@oddhead\@empty
  \let\@evenhead\@empty
  \def\@oddfoot{\small\sffamily J. Minini \& M. Wasem \hfill Page~\thepage\space of~\lastpage}
  \def\@evenfoot{\small\sffamily J. Minini \& M. Wasem \hfill Page~\thepage\space of~\lastpage}
}
}
\makeatother

\usepackage{nomencl}
\makenomenclature
\begin{document}
\let\WriteBookmarks\relax
\def\floatpagepagefraction{1}
\def\textpagefraction{.001}
\shorttitle{Structural safety}
\shortauthors{J. Minini \& M. Wasem}

\title [mode = title]{A sampling method based on highest density regions: Applications to surrogate models}                      


\author[1]{Jocelyn Minini}[type=editor,orcid=0000-0002-3813-945X]
\cormark[1]
\ead{jocelyn.minini@hefr.ch}

\credit{Conceptualization of this study, Methodology, Software}

\author[1]{Micha Wasem}[orcid=0000-0002-1310-7929]
\ead{micha.wasem@hefr.ch}

\credit{Data curation, Writing - Original draft preparation}

\affiliation[1]{organization={School of Engineering and Architecture, HES-SO University of Applied Sciences and Arts Western Switzerland},
                addressline={P\'erolles 80}, 
                postcode={1700}, 
                postcodesep={}, 
                city={Fribourg},
                country={Switzerland}}

\cortext[cor1]{Corresponding author}

\begin{abstract}
This paper introduces a practical, one-shot design-of-experiments strategy for training surrogate models in the context of uncertainty quantification. Instead of drawing the experimental design according to the input distribution, which favours training points around its mode, we propose a heuristic method to sample uniformly within a highest density region (HDR) of the input random vector, aiming at a more homogeneous approximation accuracy over the whole domain. The approach is assessed through four error metrics: two task-agnostic accuracy measures (leave-one-out and relative mean square error) and two task-oriented measures quantifying the recovery of reference global sensitivity indices and of the probability of failure.

Across the benchmark, the HDR-based designs are comparable to, or more accurate than, the distribution-based designs for most surrogate-task combinations, with the clearest gains in predictive accuracy and reliability and essentially no effect on the recovery of sensitivity indices. The method operates in a black-box context and is compatible with existing uncertainty quantification frameworks for low-dimensional and moderately correlated inputs, the curse of dimensionality affecting the sample generation rather than the design strategy itself. It is most useful whenever a single surrogate is reused across several downstream UQ tasks.
\end{abstract}


\maketitle

\section{Introduction}\label{chap_Introduction}

Most engineering and physics problems require understanding processes that can be highly complex. Within the scientific community, variational formulations are widely used to describe such problems. In this context, the prediction of a system's behaviour is typically obtained by solving a set of partial differential equations (PDEs). It is well known that such systems are often extremely difficult to solve analytically and, in most cases, do not admit easily evaluable closed-form solutions. Since the 1970s, one of the most successful numerical methods for approximating the solution of a differential problem has been the finite element method (FEM)~\citep{zienkiewiczFiniteElementMethod2025}, whose potential for handling nonlinearities was rapidly recognised by \citet{odenFiniteElementsNonlinear1972}. In addition to the difficulty of identifying an appropriate model $y = \Mcal(\xbs)$, most variables involved in the system description are subject to uncertainty. The input and output variables can therefore be represented probabilistically by random vectors. Extracting the statistical properties of these vectors requires applying a dedicated uncertainty quantification (UQ) framework~\citep{sudret2007uncertainty}, whose use often involves computationally expensive methods such as Monte Carlo (MC) simulation. Although the increased computational power of modern computing devices allows deterministic complex FE problems to be solved within reasonable time frames, combining FEM with uncertainty propagation remains computationally demanding and can be prohibitive for large-scale FE models. To address this limitation, so-called \textit{surrogate models}, \textit{metamodels}, or \textit{response surfaces} have attracted significant attention~\citep{azarhooshReviewRecentAdvances2025}.

One factor influencing the quality of a surrogate model is the sampling strategy used for the experimental design. In their review on sparse polynomial chaos expansion (PCE), \citet{luthenSparsePolynomialChaos2021a} classify sampling strategies into four categories: (i) sampling based on the input distribution (referred to here as \textit{natural sampling}); (ii) sampling from a different distribution (also called \textit{induced sampling}); (iii) optimal sampling; and (iv) adaptive sampling. Strategies (i) to (iii) are often referred to as \textit{one-shot} methods~\citep{crombecqNovelSequentialDesign2009}, where sample generation requires neither knowledge of the specific problem nor model evaluations. These contrast with the sequential approach (iv), where the adaptive algorithm generates new samples in regions that are difficult to approximate. In this work, we focus solely on strategies (i) and (ii).

The most straightforward method is natural sampling, in which samples are generated according to the probability distribution of the input random vector. In this context, several authors~\citep{blatmanQuasiRandomNumbers2007,schobiPCKrigingNewMetamodelling2014,josephSpacefillingDesignsComputer2016,santnerDesignAnalysisComputer2018a,luthenSparsePolynomialChaos2021a} have shown that \textit{space-filling} designs such as Latin Hypercube Sampling (LHS)~\citep{mckayComparisonThreeMethods1979a} or quasi-random sequences such as Sobol sequences~\citep{sobolUniformlyDistributedSequences1976} or Niederreiter sequences~\citep{niederreiterRandomNumberGeneration1992} produce samples that are particularly uniformly distributed in the $d$-dimensional unit hypercube $[0,1]^d$ and can improve the accuracy of the surrogate model. The mapping from the unit hypercube to the underlying random vector is performed using an isoprobabilistic transformation~\citep{rosenblattRemarksMultivariateTransformation1952}, 
which produces samples distributed according to the underlying distribution in the natural space even if they do not come from Monte Carlo sampling on $[0,1]^d$.

In the literature, the strategy known as sampling from a different distribution (or induced sampling) consists in modifying the probability density of the input variables and sampling according to a more favourable measure. In the context of full- and sparse PCE, the contributions of \citet{hamptonCoherenceMotivatedSampling2015,hamptonCompressiveSamplingPolynomial2015a} show that sampling designed to minimise a coherence parameter stabilises regression and reduces the mean squared error, both on manufactured stochastic functions and on a 20-dimensional elliptic PDE. In the same perspective, \citet{narayanStochasticCollocationUnstructured2015} establish stability and accuracy criteria for collocation on unstructured multivariate meshes, providing a framework to guide the selection of random or low-discrepancy grids when the input distribution is known. \citet{narayanChristoffelFunctionWeighted2016} also propose the so-called Sparse Approximation, in order to improve conditioning and reduce the number of required samples.

The contribution of this paper falls into a similar category but follows a geometric and heuristic rule. Instead of generating the experimental design according to the natural distribution of the input variables, we propose to sample it uniformly over a compact domain. Since most distributions relevant for practical applications have an unbounded support and are unimodal, we define this domain as the Highest Density Region (HDR) of probability $1-\alpha$ of the input random vector, where $\alpha$ is a tuning parameter controlling the extent of the sampling domain. The intuition behind this choice is the following: under natural sampling, training points concentrate around the mode of the underlying distribution, so that the resulting surrogate approximates the unknown model well near the mode but not necessarily elsewhere in the support. Uniform sampling within an HDR, by contrast, favours no region over another, and is therefore expected to yield more homogeneous approximation accuracy over the whole domain, including the low-density regions that natural sampling rarely visits. We emphasise that the proposed strategy operates exclusively at the level of the experimental design generation: it is non-intrusive, agnostic to the input distribution, to the surrogate type and to the downstream task, and can therefore be plugged into any existing surrogate-modelling workflow. Moreover, as all training points are generated at once, the construction of the training set is fully parallelisable, which is particularly advantageous for computationally expensive models. This generality, however, comes at a price. By design, HDR coverage spends its budget over the whole domain and cannot concentrate samples where a specific downstream task would require them; sequential designs tailored to a single, well-identified quantity of interest can therefore be expected to make more efficient use of the same evaluation budget in such settings. In addition, the practical applicability of HDR sampling is currently limited to low-to-moderate input dimensions: this restriction stems from the generation of uniform samples within an HDR, whose acceptance-rejection-based implementation is subject to the curse of dimensionality. These trade-offs are discussed throughout the study.

We apply the proposed strategy to three distinct surrogate modelling methods, namely PCE, ordinary Kriging (OK) and polynomial chaos Kriging (PCK); for polynomial-based surrogate models, the choice of the polynomial family is based on the natural input random vector, but the samples are explicitly generated according to a different distribution. The approach is evaluated on nine problems, including analytical functions, structural engineering and geomechanical problems. The predictive capabilities of the surrogates trained with the two sampling strategies are assessed through four performance metrics covering complementary aspects of surrogate quality: two task-agnostic accuracy measures, and two task-oriented measures addressing global sensitivity analysis and reliability estimation, the latter with failure probabilities ranging from $10^{-6}$ to $10^{-2}$. This is a task for which sequential, goal-oriented designs have been shown to be substantially more efficient than one-shot designs when a single failure event is of interest \citep{bichonEfficientSurrogateModels2011,moustaphaActiveLearningStructural2022a}, while one-shot designs retain a practical advantage precisely through their reusability \citep{romeroSetTestProblems2016}, e.g.\ across several reliability thresholds or alongside other downstream analyses of the same surrogate. To obtain statistically meaningful and interpretable results, the batch of computations is repeated 100 times.

This paper is organised as follows. Section \ref{chap_Methods} introduces the mathematical formulations for both PCE- and Kriging-models, presents the basis of our sampling scheme and defines the four error metrics according to which the two sampling methods, i.e.\ natural sampling and HDR sampling, are compared. Section \ref{chap_Numerical_implementation} gives a short overview of our numerical implementation for uniformly sampling an HDR and presents the 9 benchmark problems. Section \ref{chap_Results} reports the main results of the proposed methodology first with a graphical intuition, then with more detailed ranking strategy between the methods and finally with an extra $d$-dimensional toy function where the influence of the dimension combined with the probability level $\alpha$ is discussed. Conclusions and further developments are discussed in Section \ref{chap_Conclusion}.
\section{Surrogate modelling workflow}\label{chap_Methods}

Let $\Mcal$ denote a computational model which describes a physical system. Within a black-box approach, $\Mcal$ is considered as an unknown map between a $d$-variate input $\xbs = (x_1,\ldots,x_d)$ and a scalar output: 
\begin{equation}\label{equ:trueModel}
\Mcal : \xbs \in \Dcal_{\xbs} \subset \Rbb^d \mapsto y = \Mcal(\xbs) \in \Rbb.
\end{equation}

\noindent
Because the input variables of $\Mcal$ are often subject to uncertainty, they can be represented by a random vector (RV) i.e.\ a mapping $\Xbs : \Omega \to \Dcal_{\xbs} \subset \Rbb^{d}$ from the \textit{sample space} i.e.\ the set of all possible outcomes to a measurable space called the \textit{state space}, which may be \textit{discrete} or \textit{continuous}. $\Xbs =(X_{1}, \ldots, X_{d})$ can be considered as a vector whose components are random variables with marginal cumulative distribution functions (CDF) $F_{X_1},\ldots, F_{X_d}$, where $F_{X_i} : \Dcal_{x_i} \to [0,1]$. The density of $\Xbs$ can then be related to the marginal densities by a copula function $C:[0,1]^d\to[0,1]$ such that the joint CDF $F_{\Xbs} : \Dcal_{\xbs} \to [0,1]$ can be expressed as
\begin{equation}\label{equ:copula_cdf}
F_{\Xbs}(\xbs) = C(F_{X_1}(x_1), \ldots, F_{X_d}(x_d)).
\end{equation}

Numerous copula functions have been proposed in the literature to model dependence structures \citep{joeDependenceModelingCopulas2014,phoonRiskReliabilityGeotechnical2015,sepulveda-garciaUseCopulasGeotechnical2022}. While not always optimal \citep{hanComprehensiveComparisonCopula2023}, the Gaussian copula remains a practical choice to model dependencies between input parameters \citep{masoudianGeneralFrameworkCoupled2019,yanMultivariateStructuralSeismic2022} and relies on a simple formulation. This copula only depends on the component-wise defined (not necessarily Gaussian) marginals as well as a symmetric correlation matrix. As the Gaussian copula is prescribed by the problems considered later in this study, we introduce its formulation as 
\begin{equation}\label{equ:gausian_copula_cdf}
C(u_1,\ldots,u_d ; \Sigmabf) = \Phibf \; (\Phi^{-1}(u_1),\ldots,\Phi^{-1}(u_d) ; \Sigmabf),
\end{equation}

\noindent
where $\Phibf(\ubs;\Sigmabf)$ is the CDF of a $d$-variate Gaussian distribution with zero-mean and correlation matrix $\Sigmabf$ and $\Phi^{-1}$ is the inverse CDF of a univariate standard Gaussian distribution. Note that the approach described in this study is not restricted to one type of copula function. The limitations of our method are further discussed in Section \ref{sec_HDR}.

Due to the variability prescribed by $\Xbs$, the model response $Y=\Mcal(\Xbs)$ is obtained via uncertainty propagation and is therefore a random variable with unknown probability density function (PDF).

\subsection{Surrogate model formulation}\label{sec_Surrogate_model_formulation}

For most practical applications and particularly when the system is modelled via the finite element method, evaluating $\Mcal$ is computationally expensive. Consequently, performing uncertainty quantification directly on $\Mcal$ is unreasonable. It is thus often replaced by an easy-to-evaluate model, denoted $\widetilde{\Mcal}$ and called \textit{surrogate model}, in order to reduce the computational burden. Numerous surrogate modelling formulations have been developed over the years. Among them, polynomial chaos expansion and Kriging models have been widely used for uncertainty quantification problems in general \citep{kudelaRecentAdvancesApplications2022,azarhooshReviewRecentAdvances2025}.

\subsubsection{Polynomial chaos expansion}

PCE is a spectral representation of the model response with respect to a set of polynomials $\Psi_{\ibf}$ where $\ibf = (i_1,\ldots,i_d) \in \Nbb^d$.
These polynomials form an orthonormal basis with respect to the weighted $L^2$-inner product with weight function given by the probability density function $f_{\Xbs}$ of $\Xbs$ which amounts to
\[
\idotsint_{\Dcal_{\Xbs}} \Psi_\ibf(\xbs) \Psi_\jbf(\xbs) f_{\Xbs}(\xbs) \; \mathrm d\xbs = \delta_{\ibf\jbf},
\]

\noindent
so that there exist unique coefficients $\alpha_{\ibf}$ such that
\begin{equation}\label{equ:fullPCE}
Y = \Mcal(\Xbs) = \sum_{\ibf \in \Nbb^d} \alpha_{\ibf}\Psi_{\ibf}(\Xbs).
\end{equation}

Because this infinite expansion is numerically not affordable in practice, the representation must be truncated to a finite number of terms. This truncation may be achieved by selecting polynomials up to degree $\nu$. In this case, the cardinality of the set of retained multi-indices equals $p = \binom{d + \nu}{\nu}$ so that there exists an orthonormal basis $\{\Psi_{1}, \ldots, \Psi_{p}\}$ spanning the vector space of polynomials of degree at most $\nu$ and the truncated version of \cref{equ:fullPCE} might be written as 
\begin{equation}\label{equ:truncatedPCE}
Y \approx \widetilde{\Mcal}_{\text{PCE}}(\Xbs) = \sum_{k=1}^p \alpha_{k} \Psi_{k}(\Xbs) = (\Psi_1(\Xbs),\ldots, \Psi_p(\Xbs))\begin{pmatrix}\alpha_1\\ \vdots \\ \alpha_p\end{pmatrix} =: \Psibf\alphabf
\end{equation}

\noindent
and is often referred to as the \textit{total-degree} or \textit{full} PC representation of $\Mcal$ of degree $\nu$.

The calculation of the coefficients contained in $\alphabf$ can be achieved in several ways. In this paper, we follow the degree-adaptive non-intrusive approach suggested by \citet{blatmanAdaptiveSparsePolynomial2011b}. This method is called the least-angle regression (LAR) and eventually gives a \textit{sparse} PC representation of $\Mcal$ by selecting only the polynomials $\Psi_{i}$ that have the greatest influence on $Y$. For comprehensive details of the LAR algorithm, we refer the reader to \citet{efronLeastAngleRegression2004} and \citet{blatmanAdaptiveSparsePolynomial2009}.

\subsubsection{Kriging}

Kriging models assume that the model response can be approximated by a realisation of a Gaussian random process (GP) : 
\begin{equation}\label{equ:kriging}
Y \approx \widetilde{\Mcal}_{\text{K}}(\Xbs) = \boldsymbol{\beta}\transpose \fbs(\Xbs)+\sigma^2 Z(\Xbs).
\end{equation}

\noindent

Here, $\boldsymbol{\beta}\transpose \fbs(\Xbs)$ denotes the mean function (i.e., the trend), $Z$ is a zero-mean, unit-variance Gaussian process, and $\sigma^2$ is the associated process variance. The underlying probability space is characterised by a correlation function (or correlation family). If the mean function is replaced by a constant term such that $\boldsymbol{\beta}\transpose \fbs(\Xbs) = \beta_0$, the model is referred to as ordinary Kriging, which is one of the formulations used in this work. All Kriging predictions are constructed here  using multidimensional anisotropic ellipsoidal correlation functions based on the Matérn 5/2 family. We refer the reader to \citet{rasmussenGaussianProcessesMachine2005} and \citet{lataniotisUQLabUserManual2022a} for further details.
\subsubsection{Polynomial chaos Kriging}

To couple the global approximation capabilities provided by PCE with a kernel-based local correction, the mean of the Gaussian process can be replaced by a polynomial trend as proposed by \citet{schobiPCKrigingNewMetamodelling2014} which sets $\boldsymbol{\beta}\transpose \fbs(\Xbs)$ as the PC-representation of $\Mcal$. This model is referred to as polynomial chaos Kriging (PCK) and \cref{equ:kriging} rewrites
\begin{equation}\label{equ:pc-kriging}
Y \approx \widetilde{\Mcal}_{\text{PCK}}(\Xbs) = \sum_{k=1}^p \alpha_{k} \Psi_{k}(\Xbs)+\sigma^2 Z(\Xbs).
\end{equation}

\noindent
Note that we use the \textit{optimal} formulation of PCK \citep{schobiSurrogateModelsUncertainty2019} and that the PC expansion also follows the LAR formulation. Further note that whenever a Kriging model is evaluated at a realisation of the random vector, say $\xbs^{(i)}$, we denote the mean prediction of the GP as $\mu_{\widetilde{Y}}(\xbs^{(i)}) =: \widetilde{\Mcal}_{(\cdot)}(\xbs^{(i)})$ for simplicity.

\subsection{Generation of the training set}\label{sec_Training_set}

Non-intrusive surrogate models, as described in Section \ref{sec_Surrogate_model_formulation}, require only a finite number $n$ of realisations of the input random vector $\Xcal = (\xbs^{(1)},\ldots,\xbs^{(n)})$ called \textit{experimental design} (ED) and the corresponding model responses $\Ycal = (y^{(1)},\ldots,y^{(n)})$. The combination of the two is the \textit{training set} of the surrogate model.


\subsubsection{Natural sampling}\label{sec_Natural_sampling}

Most of the pseudorandom sampling strategies are able to generate points within the hypercube $[0,1]^d$ also referred to as the \textit{unit space}. To produce samples according to a random vector, so-called \textit{isoprobabilistic transforms} can be used. Given a multivariate standard uniform distribution $\Ubs \sim \Ucal([0,1]^d)$ and a user-defined random vector $\Xbs \sim f_{\Xbs}$, the isoprobabilistic transform 
\begin{equation}\label{equ:isoprobabilistic}
\Tcal : [0,1]^d \to \mathcal D_{\Xbs}, \Ubs \mapsto \Xbs = \Tcal(\Ubs)
\end{equation}

\noindent
enables sampling according to the distribution of $\Xbs$.

If the training set of $\widetilde{\Mcal}$ is generated according to $\Xbs$, we refer to this as \textit{natural sampling} since it follows the natural definition of $\Xbs$. This approach is commonly used in surrogate modelling \citep{lemaitreSpectralMethodsUncertainty2010,blatmanAdaptiveSparsePolynomial2011b,hamptonCompressiveSamplingPolynomial2015a}.

\begin{remark}
In its basic formulation, natural sampling has been introduced via MC sampling. Several authors (see Introduction) have shown that using alternative schemes within $[0,1]^d$ can improve the quality of $\widetilde{\Mcal}$. For instance, techniques such as LHS are shown to produce better PC expansions than traditional MC sampling.
\end{remark}

\subsubsection{HDR Sampling}\label{sec_HDR}

In general, distributions such as normal, log-normal, Gumbel, and Weibull are frequently used to model actions and resistances in load-bearing systems \citep{jcssJCSSPROBABILISTICMODEL2000,europeancommission.jointresearchcentre.ReliabilitybasedVerificationLimit2024}. These distributions are typically unimodal. When natural sampling is employed, the density is maximal near the mode, thereby favouring samples in this region, while it is low in the tails. Consequently, the quality of the surrogate model may be improved by using a uniform sampling density, as such a density does not favour any specific region. Below, we propose an approach to generate uniformly distributed samples while accounting for the original RV.

A \textit{credible region} of a probability distribution is defined as a subset of the sample space that contains a specified probability, say $1 - \alpha$. Since credible regions are not unique, as there are infinitely many ways to partition the sample space to achieve the same coverage probability, it is common to impose additional selection criteria, such as the following given by \citet{hyndmanComputingGraphingHighest1996}:

\begin{enumerate}
    \item[I.] The region occupies the smallest possible volume within the sample space.
    \item[II.] Every point within the region has a probability density at least as large as any point outside the region.
\end{enumerate}

\noindent
These conditions define a (\(1 - \alpha\))-\textit{highest density region}. However, it may still fail to be unique for certain probability densities. In this article, we will use the following definition, which provides a unique HDR in case it exists. A \((1-\alpha)\)-HDR is defined to be the superlevel set
\begin{equation}\label{equ:hdr_superlevel}
\Dcal_{\ell} = \{\xbs \in \Dcal_{\Xbs} \mid f_{\Xbs}(\xbs) \geq \ell \},
\end{equation}

\noindent
where the threshold \(\ell\) satisfies
\begin{equation}\label{equ:hdr_integral}
1 - \alpha = \idotsint_{\Dcal_{\ell}} f_{\Xbs}(\xbs) \, \mathrm{d}\xbs,
\end{equation}

\newcommand{\hdr}{\Dcal_{\ell}}

\noindent
with $\alpha \in [0,1]$ and \(  0\leqslant \ell\leqslant \sup\limits_{\xbs \in \Dcal_{\Xbs}}|f_{\Xbs}(\xbs)|.\)\\
\noindent
According to \cref{equ:hdr_superlevel}, it follows that $\hdr$ may be disjoint if $\Xbs$ is a multimodal distribution. Since the unknown $\ell$ is contained within the implicitly defined $d$-dimensional region $\hdr$, the direct resolution of \cref{equ:hdr_integral} given $\alpha$ is impossible in the general case. In the specific case of a $d$-variate Gaussian distribution, we obtain the closed-form solution  
\begin{equation}\label{equ:gaussian_ell}
\ell = \frac{1}{\sqrt{(2\pi)^d\det(\Cbf)}} \exp\left(-\frac{\chi^2_{d,1-\alpha}}{2}\right),
\end{equation}

\noindent
which satisfies \cref{equ:hdr_integral}, where $\Cbf$ is the covariance matrix obtained by $\Cbf = \Abf \Sigmabf \Abf$ with $\Abf = \text{diag}(\sigma_1,\ldots,\sigma_d)$ being the diagonal matrix containing the standard deviation of each Gaussian marginal and $\chi^2_{d,1-\alpha}$ is the $(1-\alpha)$-quantile of a chi-squared distribution with $d$ degrees of freedom. We refer the reader to the proof in appendix \ref{sec_appendix_gaussian}. To solve the general case for an arbitrary RV, we developed a numerical sampling-based approach based on MC-integration for the estimation of $\ell$. The implementation is detailed in Section \ref{sec_Estimation_HDR}.

Generating uniformly distributed samples in an arbitrary \(d\)-dimensional body is highly non-trivial in general as we will discuss hereinafter. To address this issue, we outline a non-exhaustive set of approaches. The first approach considers the desired points within the HDR as coming from an unknown truncated distribution with constant density. This interpretation allows the problem to be addressed iteratively using the Markov Chain Monte Carlo method. A second approach involves finding a direct mapping that transforms samples from the unit hypercube \( [0,1]^d \) to the HDR while preserving the uniformity properties of the samples. Although this method is theoretically appealing, it presents significant challenges as it relies on optimal transport theory: If the HDR is given by a simply connected domain $\Omega\subset \Rbb^d$ one would need to find a parametrization $T:(0,1)^d \to \Omega$ with the property that $|\det(\mathrm dT)|$ is constant in order to preserve uniformity -- this could for instance be achieved by solving certain Monge-Ampère equations, but already finding a parametrization without additional constraints is challenging if the geometry of $\Omega$ is intricate.

\begin{figure}[pos=t!]
\raggedright \includegraphics[scale = 0.8]{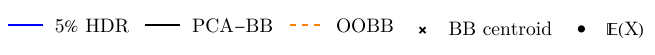}\\
\subfloat[\label{fig:Methods_BBox1} PCA]{\includegraphics[width = 0.3\linewidth]{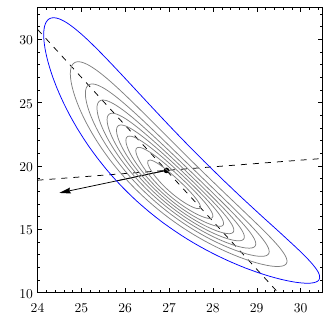}}\hfill
\subfloat[\label{fig:Methods_BBox2} Centring]{\includegraphics[width = 0.3\linewidth]{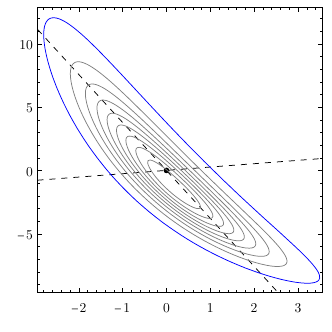}}\hfill
\subfloat[\label{fig:Methods_BBox3} Rotation]{\includegraphics[width = 0.3\linewidth]{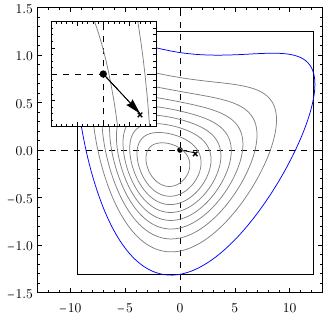}}\\
\subfloat[\label{fig:Methods_BBox4} Offset]{\includegraphics[width = 0.3\linewidth]{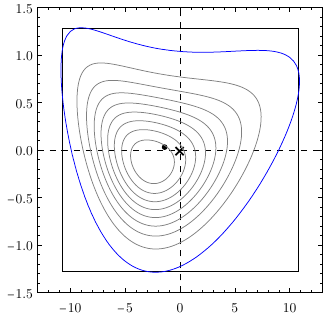}}\hfill
\subfloat[\label{fig:Methods_BBox5} Unit-space]{\includegraphics[width = 0.3\linewidth]{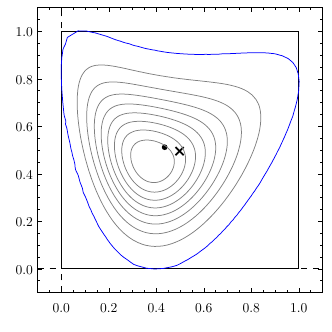}}\hfill
\subfloat[\label{fig:Methods_BBox6} Inverse transform]{\includegraphics[width = 0.3\linewidth]{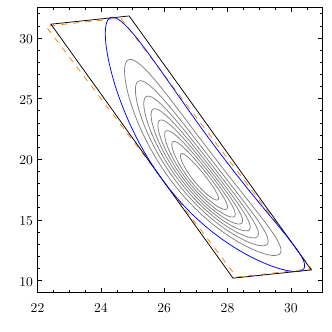}}\hfill
\caption{Affine mapping from the unit space onto the physical space. $\varphi \; (\mathrm{deg})$ on the x-axis and $c \; \mathrm{(kPa)}$ on the y-axis for (a) and (f).}
\label{fig:Methods_BBox}
\end{figure}

In this study, we adopt a simple geometric approach combined with an acceptance/rejection scheme. Standard sampling methods inherently allow for uniform sampling within \( [0,1]^d \). Therefore, our strategy is to construct an affine transformation that maps the hypercube \( [0,1]^d \) onto a bounding box that closely encloses the HDR. Samples are then filtered via an acceptance/rejection procedure, retaining only those that fall within the HDR. To illustrate the method, we provide a step-by-step two-dimensional example (\cref{fig:Methods_BBox}). We consider the study by \citet{dimatteoLaboratoryShearStrength2013}, which quantified geotechnical variability by analysing compacted alluvial fine-grained soils. This dataset consists of 256 pairs of friction angle \( \varphi \; (\mathrm{deg}) \) and cohesion \( c \; (\mathrm{kPa}) \) values, with mean values of \( \mu_{\varphi} \approx 26.9 \; (\mathrm{deg}) \) and \( \mu_{c} \approx 19.7 \; \mathrm{kPa} \). Following the approach of \citet{wangQuantifyingCrosscorrelationEffective2016}, the dependence between these variables might be modelled using a Gaussian copula with correlation matrix
\[
\Sigmabf = \begin{bmatrix} 1 & -0.92 \\ -0.92 & 1 \end{bmatrix}.
\]
The random vector representing this dataset is depicted via the probability density contours in \cref{fig:Methods_BBox1}.

In the general case, we introduce an \textit{affine mapping} from the unit cube $[0,1]^d$ into the physical space so that the image of the map encloses the HDR $\hdr$ of $\Xbs$ with level $1-\alpha$. In order to define this mapping, we will first study its inverse. Let $\Cbf\in \Rbb^{d\times d}$ be the covariance matrix of $\Xbs$, which can be orthogonally diagonalised as \[\Cbf = \Rbf \Dbf \Rbf\transpose,\]

\noindent
where $\Dbf$ is a diagonal matrix containing the eigenvalues of $\Cbf$ in \emph{descending} order and $\mathbf R\in \mathrm{SO}(d) :=\left\{\mathbf A \in \Rbb^{d\times d}\left|\mathbf A^\top\mathbf A = \mathbf{I},\det(\mathbf A) = 1\right\}\right.$ (\cref{fig:Methods_BBox1}).

In a first step, let $\xbs\mapsto\vbs(\xbs)=\Rbf(\xbs - \mathbb E(\Xbs))$. The intuition behind the map $\boldsymbol v$ is the following: First, $\boldsymbol X$ is centred by considering $\boldsymbol X-\mathbb E(\boldsymbol X)$ (see \cref{fig:Methods_BBox2}). Then we apply the rotation $\mathbf R$ onto $\Xbs-\mathbb E(\Xbs)$ in order to ensure that the principal axes of the resulting random vector are aligned with the standard coordinate axes in $\mathbb R^d$ such that the first principal axis is aligned with $(1,0,\ldots,0)^\top$, the second one with $(0,1,0,\ldots,0)^\top$ and so forth (see \cref{fig:Methods_BBox3}). Then define
\[\begin{aligned}
a_j & = \min\{v_j\in \Rbb | \exists v_i,i\in\{1,\ldots,d\}\setminus\{j\}\text{ such that }(v_1,\ldots,v_d)\in  \vbs(\mathcal D_\ell)\}\\
b_j & = \max\{v_j\in \Rbb |  \exists v_i,i\in\{1,\ldots,d\}\setminus\{j\}\text{ such that }(v_1,\ldots,v_d)\in  \vbs(\mathcal D_\ell)\}
\end{aligned}
\] for all $j=1,\ldots,d$. Here, $\vbs(\mathcal D_\ell)$ denotes the image of the map $\vbs:\mathcal D_\ell \to \Rbb^d$.
In order to explain the definition of $a_j$ and $b_j$, we will focus on $a_1$ and $b_1$. Imagine, $\boldsymbol v(\mathcal D_\ell)$ is orthogonally projected onto the first standard coordinate axis in $\mathbb R^d$. The image of this projection viewed as a map from $\boldsymbol v(\mathcal D_\ell)\to\mathbb R$ will attain a minimum and a maximum, as $\mathcal D_\ell$ is compact. The minimal value is precisely $a_1$ and the maximal value is $b_1$. Therefore, the hyper-rectangle
\[
R=\bigtimes_{j=1}^d [a_j,b_j]
\]
contains $\boldsymbol v(\mathcal D_\ell)$ and is aligned with the principal axes of $\boldsymbol v(\Xbs)$. Among all hyper-rectangles sharing these two properties, $R$ is minimal.

Upon introducing an offset vector
\[
\obs = \begin{pmatrix}\frac12(a_1+b_1)\\ \vdots \\ \frac12(a_d+b_d)\end{pmatrix},
\]
we ensure that $
|a_j-o_j| = |b_j-o_j|$ for all $j=1,\ldots, d$.
Intuitively, the offset vector $\boldsymbol o$ is introduced such that $R-\boldsymbol o$ has its center at the origin (note here that one does not necessarily have $a_j = -b_j$) see \cref{fig:Methods_BBox4}.
Now we define
$
\Sbf = \operatorname{diag}\left(b_1-a_1,\ldots,b_d-a_d\right)$
then by construction the map
\[
\xbs\mapsto \Sbf^{-1}(\Rbf(\xbs-\mathbb E(\Xbs))-\obs) + \frac12(1,\ldots,1)\transpose
\]
maps all elements in $\mathcal D_{\ell}$ onto $[0,1]^d$  (\cref{fig:Methods_BBox5}). Note that the map $\mathbf S^{-1}$ scales the shifted version of $R$ so that it becomes a hypercube with side-length $1$. This cube is finally translated such that it equals $[0,1]^d$ by adding the shift by $\frac12(1,\ldots,1)^\top$.

The inverse of the finally obtained map is given by
\begin{equation}\label{equ:linear_mapping}
T_{\ell}:\ubs \mapsto \Rbf\transpose\left[\Sbf\left(\ubs-\frac12(1,\ldots,1)\transpose\right)+\obs\right]+\mathbb E(\Xbs)
\end{equation}



\noindent
which maps given samples $(\ubs^{(1)},\ldots,\ubs^{(n)})$ living in the unit space $[0,1]^d$ into a box enclosing $\Dcal_{\ell}$ (\cref{fig:Methods_BBox6}). This box is referred to as the \textit{PCA-based bounding box} (PCA-BB) of $\Dcal_{\ell}$. A PCA-BB can in some cases be far from being the \textit{optimal oriented bounding box} (OOBB). In the present example, the PCA-BB is a fair approximation of the OOBB as shown by the orange dashed box in (\cref{fig:Methods_BBox6}). Computing the OOBB is not an easy task \citep{changFastOrientedBounding2011} already for the three-dimensional case and so is the $d$-dimensional case as well \citep{shahidFindingOptimizedBounding2014}. We discuss in the sequel why generating the OOBB is not necessary in the present case. Note that in the specific case of an independent random vector, the correlation matrix is $d$-unit and no rotation occurs in \cref{equ:linear_mapping}, the PCA-BB is then a so-called \textit{axis-oriented bounding box}.

The problem of selecting samples within $\hdr$ can be viewed by analogy with selecting those uniformly generated samples in $[0,1]^d$ which lie in the inscribed $d$-ball $B_d$ with unit-diameter within $[0,1]^d$. This mimics the selection of uniform samples in the HDR of a $d$-variate standard Gaussian random vector. Note that the probability that a sample lies in this $d$-ball equals $\operatorname{vol}_d(B_d)$. It is well known \cite{keoghCurseDimensionality2017,venkatCurseDimensionalityOut2018} that this problem suffers from the curse of dimensionality as
\begin{equation}\label{equ:nBall_volume}
\lim_{d \rightarrow \infty}\operatorname{vol}_d(B_d)= \lim_{d \rightarrow \infty}\frac{\sqrt{\pi^d}}{2^{d} \; \Gamma\left(\frac{d}{2}+1\right)} = 0.
\end{equation}

\noindent
Thus when dealing with high-dimensional independent or with highly correlated input distributions, generating samples within $\hdr$ can have a near-zero-probability and is computationally not affordable: If $d=16$ in the example above, the resulting probability is already smaller than $10^{-5}$. The generation of samples as close as possible to the HDR hence speeds up the acceptance/rejection algorithm considerably even if the PCA-BB is not the OOBB of $\Dcal_{\ell}$. In Section \ref{sec_Sampling_HDR}, we give comprehensive details on our implementation.

\subsection{Performance metrics}\label{sec_performance_metrics}

To properly compare natural and HDR sampling, some performance measures have to be defined. We propose the selection of four different metrics: the relative leave-one-out error (RLOO), the relative mean square error (RMSE), a relative sensitivity index error (RSIE) and a relative reliability index error (RRIE). For each metric, we argue for our choice in its respective section. The latter two metrics measure the deviation of a surrogate-based estimate from a deterministic reference quantity, the surrogate-based estimate being itself obtained by simulation and therefore affected by numerical noise. For each of them, we consequently derive a \textit{noise-only tolerance}, i.e.\ a threshold below which the observed error cannot be statistically distinguished from the numerical noise of the surrogate-based estimate, and below which the corresponding result is treated as having attained reference-level accuracy. Both tolerances are constructed as one-sided hypothesis tests at a probability level $\gamma = 0.01$, used throughout this study.

\subsubsection{Relative leave-one-out error}

Recall that for a given experimental design $\Xcal$, we know the exact response $\Ycal$ of the unknown model we wish to approximate by $\widetilde {\mathcal M}$. The philosophy of the leave-one-out error \citep{stoneCrossvalidatoryChoiceAssessment1974} consists of removing a point -- say $\bm x^{(i)}$ -- from the experimental design and creating a metamodel $\widetilde{\mathcal M}^{\hat\imath}$ based on the remaining points. The RLOO is then defined as
\begin{equation}\label{equ:error_loo}
\varepsilon_{\text{RLOO}} = \dfrac{\frac{1}{n}\sum_{i=1}^n (y^{(i)}-\widetilde{\mathcal M}^{{\hat\imath}}(\bm x^{(i)}))^2}{\Vbb(\Ycal)},
\end{equation}

\noindent
where $\Vbb(\cdot)$ denotes the discrete variance. Note that this definition requires a priori the creation of $n$ metamodels in order to be computed. \citet[Appendix D]{blatmanAdaptiveSparsePolynomial2009} and \citet{dubruleCrossValidationKriging1983} however give analytical formulations of \cref{equ:error_loo} for PCE and Kriging models respectively without computing $n$ surrogate models explicitly. Note that in the case of a PCE approximation, we use the corrected error estimate after \citet{chapelleModelSelectionSmall2002} to avoid underestimation of the error for small experimental designs.
Several authors \citep{molinaroPredictionErrorEstimation2005,huComparativeStudiesError2018} have shown that RLOO introduces little bias and is suitable in most cases for assessing the quality of a predictor. However, we argue that evaluating the quality of the surrogate model based solely on RLOO is insufficient and we therefore follow the conclusions drawn in a slightly different context by \citet{poldrackEstablishmentBestPractices2020} stating that multiple measures of prediction accuracy should be examined and reported.


\subsubsection{Relative mean square error}

Given an additional experimental design $\Xcal^{+}$ composed of $m$ realisations of the RV and generated so that $\Xcal \cap \Xcal^{+} = \varnothing$, the set composed of $\Xcal^{+}$ and the corresponding model responses $\Ycal^{+}$ forms the \textit{validation set} of $\widetilde{\Mcal}$. The relative mean square error is the discrete relative $L^2$-norm defined as 
\begin{equation}\label{equ:error_rmse}
\varepsilon_{\text{RMSE}} = \dfrac{\frac{1}{m}\sum\limits_{i=1}^m (y^{(i)}-\widetilde{\mathcal M}(\bm x^{(i)}))^2}{\Vbb(\Ycal^{+})},
\end{equation}

\noindent
where $\xbs^{(i)} \in \Xcal^{+}$ and $y^{(i)} \in \Ycal^{+}$.

\subsubsection{Relative sensitivity index error}

A common downstream use of surrogate models is global sensitivity analysis. We therefore introduce the \textit{relative sensitivity index error}, which measures the deviation of the surrogate-based total-effect indices $\widetilde{S}_{T,i}$ from reference values $S_{T,i,\text{ref}}$ in a relative $L^1$-sense:
\begin{equation}\label{equ:error_rsie}
\varepsilon_{\mathrm{RSIE}} = \dfrac{\sum_{i=1}^{d}\left|\widetilde{S}_{T,i} - S_{T,i,\text{ref}}\right|}{\sum_{i=1}^{d}\left|S_{T,i,\text{ref}}\right|},
\end{equation}
\noindent
where the total-effect indices are Sobol' indices for independent inputs and Kucherenko indices for correlated ones, both estimated by a pick-freeze Monte Carlo scheme.
The reference indices $S_{T,i,\text{ref}}$ are computed once to a controlled accuracy and subsequently held fixed, playing the role of deterministic reference values. The computation of $\varepsilon_{\mathrm{RSIE}}$ may then suffer from two sources of error: the error from the approximation of the original function and the one from the numerical integration itself.
These estimators provide no per-analysis convergence criterion, so the convergence of the reference is controlled empirically: for a given pick-freeze sample size $n$, the estimation is replicated $K$ times on independent samples, and $n$ is increased until the one-sided $(1-\gamma)$ Gaussian bound on the inter-replication dispersion of every total index falls below a prescribed tolerance. This Gaussian assumption rests on the asymptotic normality of the pick-freeze estimators, established for Sobol' indices by \citet{janonAsymptoticNormalityEfficiency2014} and carrying over to the Kucherenko estimators. The reference $S_{T,i,\text{ref}}$ is the mean over the $K$ replications and $\sigma_{T,i}$ the corresponding standard deviation; as this adaptive procedure is computationally unaffordable for every surrogate of the benchmark, each $\widetilde{S}_{T,i}$ is obtained from a single pick-freeze run at the converged $n$. The observed error $\varepsilon_{\mathrm{RSIE}}$ stems from two sources: the surrogate approximation, which we wish to quantify, and the Monte Carlo noise of the pick-freeze estimator. The noise-only tolerance $\bar{\varepsilon}_{\mathrm{RSIE}}$ characterises the second alone: it is the value that $\varepsilon_{\mathrm{RSIE}}$ would not exceed, with probability $1-\gamma$, were the surrogate exact. In that case each deviation $\widetilde{S}_{T,i} - S_{T,i,\text{ref}}$ reduces to the pick-freeze noise of $\widetilde{S}_{T,i}$, asymptotically Gaussian with zero mean and standard deviation $\sigma_{T,i}$, so that its absolute value follows a folded normal distribution. Approximating the sum $\sum_{i=1}^{d}|\widetilde{S}_{T,i} - S_{T,i,\text{ref}}|$ by a Gaussian with matching mean and variance, the $(1-\gamma)$-quantile of $\varepsilon_{\mathrm{RSIE}}$ is given by
\begin{equation}\label{equ:noise_tol_rsie}
\bar{\varepsilon}_{\mathrm{RSIE}} := \dfrac{\sqrt{2/\pi}\,\sum_{i=1}^{d}\sigma_{T,i} \;+\; z_{1-\gamma}\,\sqrt{\left(1-2/\pi\right)\sum_{i=1}^{d}\sigma_{T,i}^{2}}}{\sum_{i=1}^{d}\left|S_{T,i,\text{ref}}\right|}.
\end{equation}

\begin{remark}
For PCE surrogates, Sobol' indices can be obtained analytically by post-processing the expansion coefficients, at virtually no computational cost \citep{sudretGlobalSensitivityAnalysis2008}. To our knowledge, no analogous post-processing is established for the Kucherenko indices. Since the present study compares three surrogate types, only one of which admits such a shortcut, all sensitivity indices are estimated by the same sample-based Monte Carlo scheme, applied uniformly to every surrogate for consistency.
\end{remark}

\subsubsection{Relative reliability index error}

A metric of particular interest to the practitioner is the ability of the surrogate model to correctly estimate the probability of failure of the system described by $\Mcal$. We thus introduce the \textit{relative reliability index error}, which measures the deviation of the reliability index $\widetilde{\beta}$ predicted by the surrogate $\widetilde{\Mcal}$ from a reference value $\beta_{\text{ref}} > 0$.
\begin{equation}\label{equ:relative_beta_index}
\varepsilon_{\text{RRIE}} = \dfrac{\left|\widetilde{\beta}-\beta_{\text{ref}}\right|}{\beta_{\text{ref}}}.
\end{equation}

In general settings, a reliability index is derived from the probability of failure $\beta = -\Phi^{-1}(P_{f})$. Introducing the \textit{limit state function} $g(\xbs)$, which defines the failure domain $\Dcal_{f} = \{\xbs \in \Dcal_{\Xbs} | g(\xbs) \leq 0 \}$, the probability of failure can be expressed as 
\begin{equation}\label{equ:pf}
P_{f} =  \idotsint_{\Dcal_{f}} f_{\Xbs}(\xbs) \, \mathrm{d}\xbs.
\end{equation}

\noindent
It is noteworthy that this integral is analogous to that presented in \cref{equ:hdr_integral}. However, in this case, the domain of integration is entirely driven by the underlying model $\Mcal$ and more specifically via its corresponding hypersurface $g(\xbs)=0$, as typically, $g$ is just a vertical shift of $\mathcal M$.

The challenge posed by the numerical resolution of this integral has attracted much effort, resulting in a rich variety of methods \cite{jiGeotechnicalReliabilityAnalysis2023,lemaireStructuralReliability2009,morioSurveyRareEvent2014}. Among them, we have selected the \textit{importance sampling} (IS) approach for estimating $P_f$ \citep{liuEfficientSurrogateaidedImportance2019,melchersImportanceSamplingStructural1989}. IS is used as a variance reduction technique which enables sampling from a different distribution $f_{\Ibs}$ instead of sampling from $f_{\Xbs}$ thereby reducing the number of model evaluations required to estimate $P_f$. As failure probabilities are here systematically estimated via numerical methods, the estimated probability of failure, conventionally denoted $\widehat{P}_{f}$, will be denoted by $P_f:=\widehat{P}_{f}$ as well as $\beta:=\widehat{\beta}$ for simplicity. Given a large number of samples $\xbs^{(1)},\ldots,\xbs^{(N)}$ drawn from $f_{\Ibs}$, the probability of failure estimated by simulation reads
\begin{equation}\label{equ:importance_sampling}
P_f =  \dfrac{1}{N}\sum_{k=1}^{N}\onebf_{\Dcal_{f}}(\xbs^{(k)})\dfrac{f_{\Xbs}(\xbs^{(k)})}{f_{\Ibs}(\xbs^{(k)})},
\end{equation}

\noindent
where $\onebf_{\Dcal_{f}}$ is the indicator function of the failure domain. As a particular case of Monte Carlo simulation, IS provides an exact error estimate for \cref{equ:importance_sampling} with an associated coefficient of variation $\text{CoV}$. In this study, all reliability analyses guarantee $\text{CoV} < 0.001$ when estimating $P_f$. We refer the reader to Section \ref{sec_Benchmark_procedure} for implementation details.

The reference value $\beta_{\text{ref}}$ is computed once to a controlled accuracy and subsequently held fixed, playing the role of a deterministic reference value. The observed error $\varepsilon_{\mathrm{RRIE}}$ stems, as previously stated, from two sources: the surrogate approximation, which we wish to quantify, and the Monte Carlo noise of the importance-sampling estimate. The noise-only tolerance $\bar{\varepsilon}_{\mathrm{RRIE}}$ characterises the second alone: it is the value that $\varepsilon_{\mathrm{RRIE}}$ would not exceed, with probability $1-\gamma$, were the surrogate exact. An observed error below it is attributed to numerical noise, and the corresponding result is considered to have reached reference-level accuracy.

Recall that the reliability index is related to the failure probability via $\beta = -\Phi^{-1}(P_f)$, where $\Phi$ and $\varphi$ denote the standard normal CDF and PDF, respectively. Given that the target coefficient of variation is small ($c := \text{CoV} = 10^{-3}$), a first-order linearisation of the map $P_f \mapsto -\Phi^{-1}(P_f)$ around $P_f$ is reasonable. The standard deviation of the surrogate-based estimate $\widetilde{\beta}$ is thus approximated by
\begin{equation}
\sigma_{\beta} \approx \frac{c \, \Phi(-\beta)}{\varphi(\beta)}.
\end{equation}

\noindent
were the surrogate exact, $\widetilde{\beta} - \beta_{\text{ref}}$ would reduce to the importance-sampling noise of $\widetilde{\beta}$, asymptotically Gaussian with zero mean and standard deviation $\sigma_\beta$, so that $\varepsilon_{\mathrm{RRIE}}$ follows a folded normal distribution. Its $(1-\gamma)$-quantile, evaluated at $\beta_{\text{ref}}$, defines the noise-only tolerance:
\begin{equation}\label{equ:noise_tol_rrie}
\bar{\varepsilon}_{\mathrm{RRIE}} := z_{1-\gamma/2} \; \frac{c \, \Phi(-\beta_{\text{ref}})}{\beta_{\text{ref}} \, \varphi(\beta_{\text{ref}})}.
\end{equation}
\section{Numerical implementation}\label{chap_Numerical_implementation}

To propose an approach compatible with existing uncertainty quantification frameworks, we used the \matlab-based \uqlab library \citep{marelliUQLabFrameworkUncertainty2014}. This library provides extensive functionality for surrogate modelling and uncertainty quantification and has undergone rigorous validation. Our approach integrates into \uqlab 's existing workflow and comprises two primary functions: estimation of the HDR and generation of uniformly distributed points in the latter.

\subsection{Algorithms}\label{sec_Algorithms}

\subsubsection{Estimation of the HDR}\label{sec_Estimation_HDR}

\begin{figure}[pos=t!]
\subfloat[\label{fig:drawHDR1} Gaussian]{\includegraphics[scale=1]{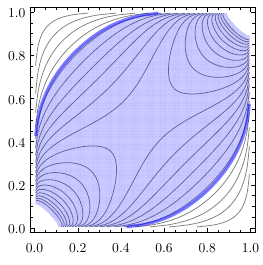}}\hfill
\subfloat[\label{fig:drawHDR2} Frank]{\includegraphics[scale=1]{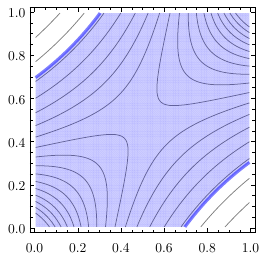}}\hfill
\subfloat[\label{fig:drawHDR3} Clayton]{\includegraphics[scale=1]{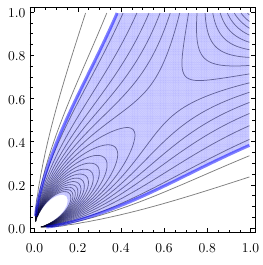}}\hfill
\subfloat[\label{fig:drawHDR4} t]{\includegraphics[scale=1]{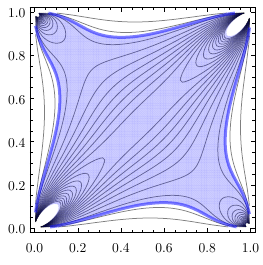}}
\caption{Contours of the density function for uniform random vectors $\Ucal(0,1)$ with several copula families and the corresponding 5\% HDR.}
\label{fig:drawHDR}
\end{figure}

The global scheme of our implementation is briefly described in \cref{alg:createHDR}. In the very special case where the random vector has constant (e.g. when all marginals are uniform and independent) or piecewise constant density, the algorithm returns nothing for $\ell$ and the HDR is defined as the original compact support prescribed directly by the marginals. In the general case and as outlined in \cref{equ:hdr_integral}, the HDR estimation can be reformulated as a standard reliability problem where the failure domain is represented by the complement of $\Dcal_{\ell}$. For this purpose, we employed the reliability module of UQL\scriptsize ab \normalsize \citep{marelliUQLabUserManual2022} in combination with a root-finding algorithm. The PCA-BB and its corresponding matrices are then approximated based on Monte-Carlo sampling with a large number of samples. It is worth noting that even for very simple marginal distributions, the hypersurface of the HDR $\partial \Dcal_\ell$ can be far from being trivial. As a minimal example, \cref{fig:drawHDR} shows the density contours of four standard uniform random vectors $\Ucal(0,1)^2$ described with four different copula families. In some cases, we note that the corresponding 5\% HDR given in blue may be non-convex which highlights the difficulty of finding a general sampling method which enables uniform sampling within $\Dcal_\ell$.

\begin{algorithm}[b]
\caption{Highest Density Region (HDR)}
\label{alg:createHDR}
\begin{algorithmic}[1]
\State \textbf{Input:} Input distribution $f_{\Xbs}$ , probability level $\alpha \in (0,1)$, tolerance $\text{tol} > 0$
\State \textbf{Output:} Unknown $\ell$ and transformation matrices $\Mbf$, $\Rbf$ and $\Sbf$. 

\State $\Xcal \leftarrow \textsc{Sample}(f_{\Xbs}, n)$ \Comment{Sampling with large $n$}

\If{$f_{\Xbs}$ has constant density}
    \State $\ell \leftarrow \text{NaN}$
    \State $\Xcal_\alpha \leftarrow  \Xcal$
\Else

    \State $\Fcal \leftarrow f_{\Xbs}(\Xcal)$
    \State $\xbs_0 \leftarrow \textsc{Quantile}(\Fcal, \alpha)$ \Comment{Initial value}
    
    \State Define $g(\ell) = \textsc{LevelIntegrate}(\ell, f_{\Xbs}, \text{tol}) - \alpha$ \Comment{Reliability problem using \uqlab}
    \State $\ell \leftarrow \textsc{FindRoot}(g, \xbs_0 )$ \Comment{Root search with fzero}
    
    \State $\Xcal_\alpha \leftarrow \{\xbs\in \Xcal : f_{\Xbs}(\xbs) > \ell\}$
\EndIf

\State $[\mathbf{R}, \mathbf{S}, \mathbf{M}] \leftarrow \textsc{GetTransform}(f_{\Xbs}, \Xcal_\alpha)$ \Comment{Transformation matrices of the PCA-BB}

\State \Return 
\State \quad $\bullet$ $\ell$
\State \quad $\bullet$ $\mathbf{S}$, $\mathbf{R}$ and $\mathbf{M}$
\end{algorithmic}
\end{algorithm}

Our implementation was validated using the closed-form solution for $\ell$ given in \cref{equ:gaussian_ell}. The Gaussian random vector $\Xbs \sim \Ncal(\mathbf{0},\Cbf)$ was randomly generated with the following properties:

\begin{enumerate} \item[I.] All marginals have zero mean. \item[II.] All marginal standard deviations satisfy $\sigma_i \sim \Ucal(0,20)$ for $i=1,\ldots,d$. \item[III.] The correlation matrix $\Sigmabf$ was randomly generated 
using the randcorr routine within the gallery function in \matlab.
\end{enumerate}

\begin{figure}[pos=t!]
\subfloat[\label{fig:Benchmark_ValidationHDR1} $\alpha = 0.01$ \& $\text{CoV} = 0.01$]{\includegraphics[width = 0.24\linewidth]{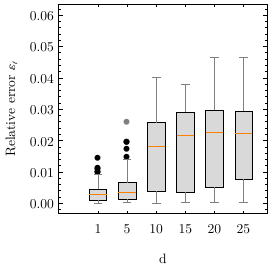}}\hfill
\subfloat[\label{fig:Benchmark_ValidationHDR2} $d = 10$ \& $\text{CoV} = 0.01$]{\includegraphics[width = 0.24\linewidth]{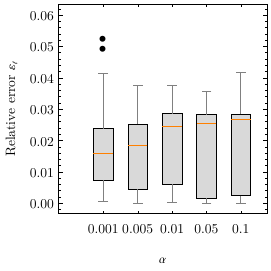}}\hfill
\subfloat[\label{fig:Benchmark_ValidationHDR3} $d = 10$ \& $\alpha = 0.01$]{\includegraphics[width = 0.24\linewidth]{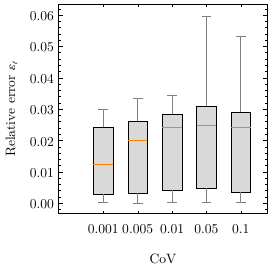}}\hfill
\subfloat[\label{fig:Benchmark_ValidationHDR4} $\ell$ vs $\widehat{\ell}$]{\includegraphics[width = 0.24\linewidth]{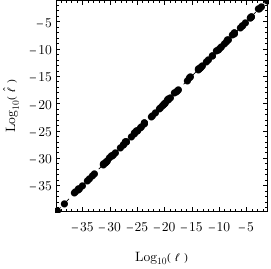}}
\caption{Sensitivity analysis on the error when estimating $\ell$.}
\label{fig:Benchmark_ValidationHDR}
\end{figure}

Using 100 random replications, we investigated the influence of the dimension $d$, the HDR probability $\alpha$, and the target CoV on the relative error, defined as
\begin{equation}\label{equ:relative_ell} 
\varepsilon_{\ell} = \dfrac{|\widehat{\ell}-\ell|}{\ell}, 
\end{equation}

\noindent where $\ell$ is the level set computed using the closed-form \cref{equ:gaussian_ell}, and $\widehat{\ell}$ is its estimated value from our implementation. The results of this analysis are presented in \cref{fig:Benchmark_ValidationHDR1,fig:Benchmark_ValidationHDR2,fig:Benchmark_ValidationHDR3}. In a subsequent step, we performed 100 additional random verifications, setting $d \sim \Ucal(1,25)$, $\alpha \sim \Ucal(0.001,0.1)$, and CoV$\sim \Ucal(0.001,0.1)$ which are shown in the parity plot in \cref{fig:Benchmark_ValidationHDR4}.

None of the 100 replications of the sensitivity analyses carried out on $d$, $\alpha$ and CoV showed a $\ell$-relative error of more than 6\%. A few trends can nevertheless be observed: the sensitivity analysis on the dimension shows that the error increases with the dimension of the random vector; for a fixed dimension $d=10$ and $\text{CoV} = 0.01$, the error decreases when the probability level increases; the error made on $\ell$ increases when the CoV of the MC simulation increases during the approximation of the integral \eqref{equ:hdr_integral}. Note that since Monte Carlo simulation is involved, the target accuracy, defined by the CoV of the MC estimate, can be arbitrarily set by the user depending on the available computing capabilities.

\subsubsection{Uniformly sampling an HDR}\label{sec_Sampling_HDR}

\begin{figure}[pos=b!]
\subfloat[\label{fig:SamplingHDR4}$\Sigmabf = \begin{bmatrix} 1 & 0 \\ 0 & 1 \end{bmatrix}$]{\includegraphics[scale=1]{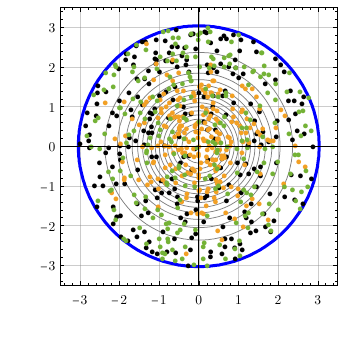}}\hfill
\subfloat[\label{fig:SamplingHDR5}$\Sigmabf = \begin{bmatrix} 1 & 0.77 \\ 0.77 & 1 \end{bmatrix}$]{\includegraphics[scale=1]{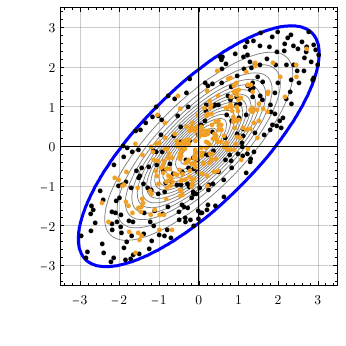}}\hfill
\subfloat[\label{fig:SamplingHDR6}$\Sigmabf = \begin{bmatrix} 1 & 0.32 & 0.02 \\ 0.32 & 1 & -0.2 \\ 0.02 & -0.2 & 1 \end{bmatrix}$]{\raisebox{0.5cm}{\includegraphics[scale=1]{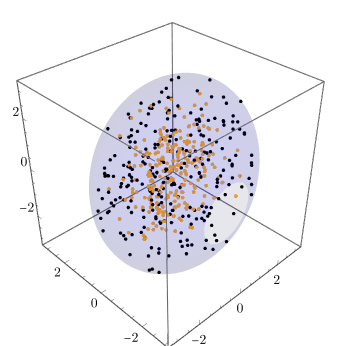}}}
\caption{Uniformity of the HDR sampling against a filtered Gaussian random vector. (a) to (c) scatters of the samples within the HDR. 
}
\label{fig:SamplingHDR}
\end{figure}

\begin{algorithm}[t!]
\caption{HDR Sampling}
\label{alg:hdr_sampling}
\begin{algorithmic}[1]
\State \textbf{Input:} Number of samples $n$, input distribution $f_{\Xbs}$, level $\ell$ and transformation function $T_{\ell}$
\State \textbf{Output:} List of samples that satisfy $\Xcal_\ell \in \Dcal_\ell^n \subset \mathbb{R}^{d \times n}$
\State $d \leftarrow \dim(f_{\Xbs})$ \Comment{Input dimension}
\State $\displaystyle n_{\text{current}} \leftarrow \Bigg\lceil  \frac{\Gamma(\frac{d}{2}+1)}{\pi^{\frac{d}{2}} \cdot \left(\frac{1}{2}\right)^d} \cdot n \Bigg\rceil$ \Comment{Initial guess based on the hyperball}
\While{$true$}

    \State $\Xcal_{\ubs} \leftarrow \textsc{Sample}(\Ucal(0,1)^d, n_{\text{current}})$ 
    \State $\Xcal_{\text{PCA-BB}} \leftarrow T_{\ell}(\Xcal_{\ubs})$                             \Comment{Linear map}
    \State $\Xcal_\alpha \leftarrow \{\xbs\in \Xcal_{\text{PCA-BB}} : f_{\Xbs}(\xbs) > \ell\}$  \Comment{Reject samples outside $\Dcal_\ell$}
    \State $score = \text{card}(\Xcal_\alpha)$
    \If{$score \geq n$}
        \State $\Xcal_\ell \leftarrow  \textsc{Subsample}(\Xcal_\alpha, n)$               \Comment{Return exactly $n$ samples}
        \State \textbf{break}
    \Else
        \State $n_{\text{current}} \leftarrow n_{\text{current}}^{++}$                     \Comment{Increase number of samples}
    \EndIf
\EndWhile

\State \Return  $\Xcal_\ell$
\end{algorithmic}
\end{algorithm}

As stated previously, the sampling method adopted in this study to sample inside the HDR is an acceptance–rejection scheme. As summarised in \cref{alg:hdr_sampling}, samples are generated after the HDR has been approximated, because both the linear transformation $T_\ell$ and the level $\ell$ are required. To ensure reasonable convergence speed, the initial sample size can be set equal to the expected number of realisations needed to obtain $n$ accepted points in the unit hyperball. The sample size is then successively increased until the number of points in the HDR equals or exceeds the target value; in the latter case, a subsampling step is applied.




For any random vector, the uniformity of the samples inside the HDR follows immediately from the fact that $T_\ell$ is an invertible affine transformation and that restrictions of uniform random vectors to full-dimensional subsets preserve uniformity. \cref{fig:SamplingHDR} illustrates \cref{alg:hdr_sampling} for the particular case of a Gaussian random vector. For three distinct correlation matrices, \cref{fig:SamplingHDR4,fig:SamplingHDR5,fig:SamplingHDR6} show (i-black) the $n=250$ points generated by the acceptance–rejection algorithm of \cref{alg:hdr_sampling}, and (ii-orange) realisations of a standard Gaussian vector filtered by the condition $f_{\Xbs}(\xbs)>\ell$, with $n=250$ likewise enforced. When the correlation is zero, the HDR reduces to a hyperball in which samples can be drawn uniformly using the method of \citet{mullerNoteMethodGenerating1959} also used by \citet[Eq.(29)]{hamptonCoherenceMotivatedSampling2015}.  For comparison, this method is shown in \cref{fig:SamplingHDR4} (iii-green). 

\subsection{Case studies}\label{sec_case_studies}

As the proposed sampling scheme is purely heuristic, we propose its validation through testing. Nine problems have been selected and are detailed in \cref{tab:problemsList}. For the reference sensitivity indices of each model, we refer the reader to \cref{tab:refSensitivity} in Appendix \ref{sec_appendix_refsensitivity}. This study focuses only on analytical, structural and geotechnical models to validate the approach. Two models require the finite element method, one of which represents a practical real case of geotechnical work in Switzerland. For the unreferenced models, a comprehensive description is provided in the sequel. For the others, the reader is invited to refer to the studies cited in \cref{tab:problemsList}.

\begin{table}[pos=h]
\centering
\small
\caption{Overview of the 9 problems used in our study. Finite element models are subscripted with $_{\text{(FE)}}$, the others have analytical formulations. Problems marked with $^{(\ast)}$ are not reproducible.}
\label{tab:problemsList}
\begin{tabular}{cllccccc}
\toprule
Acronym & Name & Problem type & Dimension  & Correlation & \multicolumn{2}{c}{Reference reliability} & Reference \\
& & & & ? & $\beta_{\text{ref}}$ & $P_{f,\text{ref}}$ & \\
\midrule
$\Mcal_1$ & \hyperref[sec_franke]{Franke function} & Analytical  & 2 & $\times$  & $1.58$ & $5.74 \times 10^{-2}$ & This study \\
$\Mcal_2$ & Short column                  & Structural & 3 & $\checkmark$ & $2.51$ & $5.97 \times 10^{-3}$  & \cite{kirjner-netoAlgorithmsReliabilityBasedOptimal1995}  \\
$\Mcal_3$ & \hyperref[sec_strip_foot]{Strip foundation} & Geotechnical & 4 & $\checkmark$ & $3.47$ & $2.57 \times 10^{-4}$ & This study \\
$\Mcal_4$ & Bracket structure             & Structural  & 5 & $\times$  & $2.00$ & $2.29 \times 10^{-2}$ & \cite{pannerselvamScarceSampleBasedReliability2022} \\
$\Mcal_5$ & Infinite slope                & Geotechnical & 6 & $\times$  & $1.58$ & $5.76 \times 10^{-2}$  & \cite{phoonReliabilityBasedDesignGeotechnical2008} \\
$\Mcal_6$ & \hyperref[sec_sheet_pile_wall]{Sheet pile wall} & Geotechnical$^{(\ast)}_{\text{(FE)}}$ & 7 & $\checkmark$ & $3.07$ & $1.07 \times 10^{-3}$  & This study \\
$\Mcal_7$ & Steel column                  & Structural & 9 & $\times$  & $4.11$ & $1.94 \times 10^{-5}$  & \cite{kuschelTwoBasicProblems1997}  \\
$\Mcal_8$ & Truss structure               & Structural$_{\text{(FE)}}$ & 10 & $\times$ & $2.96$ & $1.52 \times 10^{-3}$ &\cite{schobiSurrogateModelsUncertainty2019}\\
$\Mcal_9$ & $d$-dimensional & Analytical & $d$ & $\times$ & \multicolumn{2}{c}{User defined} & This study \\
\bottomrule
\end{tabular}
\end{table}

\subsubsection{Franke function $\Mcal_1$}\label{sec_franke}

The Franke function \citep{frankeCriticalComparisonMethods1979} is an exponential function sometimes used in regression and interpolation problems \citep{hamptonBasisAdaptiveSample2018}. The equation is given by
\begin{align}
\Mcal_1(\Xbs) = & 
\frac{3}{4} \exp\left(-\frac{(9X_1 - 2)^2}{4} - \frac{(9X_2 - 2)^2}{4}\right) + 
\frac{3}{4} \exp\left(-\frac{(9X_1 + 1)^2}{49} - \frac{(9X_2 + 1)}{10}\right) + \nonumber\\
&+ \frac{1}{2} \exp\left(-\frac{(9X_1 - 7)^2}{4} - \frac{(9X_2 - 3)^2}{4}\right) - \frac{1}{5} \exp\left(-(9X_1 - 4)^2 - (9X_2 - 7)^2\right).
\end{align}

\begin{figure}[pos=t!]
\raggedright \includegraphics[scale = 0.8]{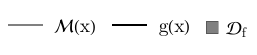}\\
\subfloat[\label{fig:Benchmark_FrankeSketch}]{\includegraphics[scale=1]{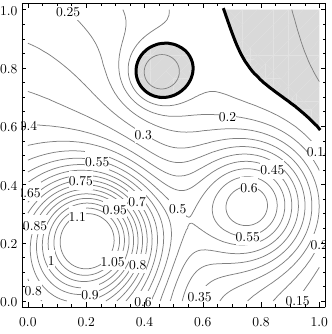}}\hfill
\subfloat[\label{fig:Benchmark_d_dimensional2}]{\includegraphics[scale=1]{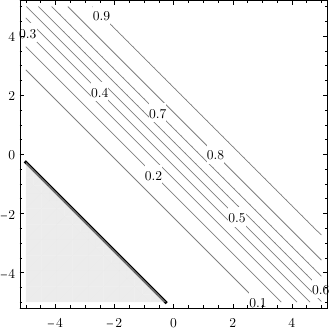}}\hfill
\subfloat[\label{fig:Benchmark_d_dimensional1}]{\includegraphics[scale=1]{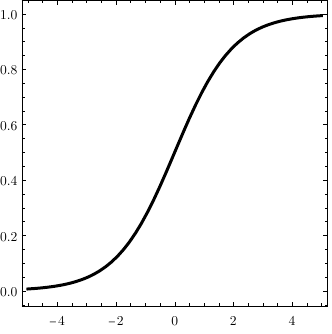}}\hfill
\caption{(a) Contour plot of $\Mcal_1$ with its corresponding $g$-function. (b) Contour plot of $\Mcal_9$ with $d=2$ and $P_f = 10^{-4}$. (c) Hyperbolic perturbation function $h$.}
\label{fig:Benchmark_Franke_d_dimensional}
\end{figure}

\noindent
\cref{fig:Benchmark_FrankeSketch} shows some contour lines of the Franke function. To highlight our approach, we impose the input variables to follow a bivariate Gaussian distribution with independent marginals $X \sim \Ncal(0.5,0.2)^2$. We furthermore define the limit-state function as $g_1(\xbs) = \Mcal_1(\xbs) - 0.1$ also shown in \cref{fig:Benchmark_FrankeSketch}.

\subsubsection{d-Dimensional function $\Mcal_9$}\label{sec_d_dimensional}

To enable more advanced analyses of the influence of the dimension $d$ and the HDR probability level $\alpha$, we introduce the function
\begin{equation}\label{equ:d_dimensional}
\Mcal_9(\Xbs) = h \left( \sum_{i=1}^d X_i \right) - h \left( \sqrt{d} \, \Phi^{-1}(P_f) \right).
\end{equation}

\noindent
If $h : \Rbb \to \Rbb$ is a strictly increasing injective function and $f_{\Xbs} \sim \Ncal(0,1)^d$, then $P_f$ can be prescribed a priori while ensuring that
\[
P_f = \idotsint_{\Dcal_f} f_{\Xbs}(\xbs)\, \mathrm{d}\xbs,
\]

\noindent
where $\Dcal_f = \{ \xbs \in \Dcal_{\Xbs} \mid \Mcal_9(\xbs)<0 \}$ is the failure domain represented in \cref{fig:Benchmark_d_dimensional2}. The reader is referred to Appendix~\ref{sec_appendix_d_dimensional} for further details. One notable property of this function is that it allows arbitrary choices of $h$ while retaining an exact solution for $P_f$ regardless of the chosen dimension. In this study, we selected a hyperbolic perturbation function $h : x \in \Rbb \mapsto h(x) = {1}/{(1 + \mathrm e^{-x})} \in (0,1)$ illustrated in \cref{fig:Benchmark_d_dimensional1}.

\begin{figure}[pos=b,width=.5\textwidth]
\centering
\captionbox{Sketch of $\Mcal_3$\label{fig:Benchmark_stripFootSketch}}[.5\textwidth]{\includegraphics[width=.8\linewidth]{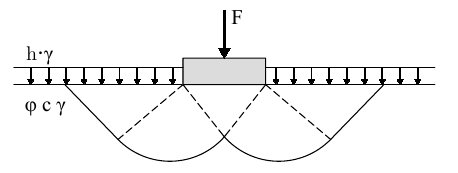}}%
\captionsetup{type=table}\captionbox{Marginal description of the probabilistic model $\Mcal_3$\label{tab:Benchmark_stripFoot}}[.5\textwidth]{\small
\begin{tabular}{ccccc}
\toprule
Variable & Unit &  $f_ {\Xbs}$ & $\mu_{\Xbs}$ & $\sigma_{\Xbs}$ \\
\midrule
$\varphi$ & $\mathrm{(deg)}$      & $\mathcal{LN}$ & 26.9 & 1.3 \\
$c$       & $\mathrm{(kPa)}$      & $\mathcal{LN}$ & 19.7 & 4.9 \\
$\gamma$  & $\mathrm{(kNm^{-3})}$ & $\mathcal{LN}$ & 21   & 1.7\\
$F$       & $\mathrm{(kN)}$       & $\mathcal{G}$  & 1400 & 140\\
\bottomrule
\end{tabular}}
\captionsetup{type=figure}
\end{figure}

\subsubsection{Strip foundation $\Mcal_3$}\label{sec_strip_foot}
The third problem represents the bearing capacity of a vertically loaded shallow strip foundation (\cref{fig:Benchmark_stripFootSketch}). The safety factor is expressed by the ratio
\begin{equation}\label{equ:stripFoot_FS}
\Mcal_3(\Xbs) = \frac{Q_p}{F}
\end{equation}

\noindent where $\Xbs = (\varphi,c,\gamma,F) \in \Rbb^4$, $F$ is the applied force and $Q_p$ is the strength of the foundation which was first defined by \citet{terzaghiTheoreticalSoilMechanics1943} and is expressed as
\begin{equation}\label{equ:stripFoot_Qp}
Q_p = b \; \left( c \; N_c + h \; \gamma \; N_q + 0.5 b \; \gamma \; N_\gamma \right),
\end{equation}

\noindent
where $\varphi$ is the friction angle, $c$ the cohesion, $\gamma$ the unit weight of the soil, $h$ the height of the overburden soil and $b$ the width of the foundation. We refer the reader to \citet{meyerhofRecentResearchBearing1963} for the formulation of the first two bearing capacity factors ($ N_c $ and $ N_q $) and to \citet{hansenRevisedExtendedFormula1970} for $ N_y $. The height of the overburden soil and the width of the strip are assumed to be deterministic with respective values $h = 0.5$ and $b = 2.7$. The marginal description of the random vector $\Xbs$ is given in \cref{tab:Benchmark_stripFoot}. A pair correlation of -0.92 and modelled via a Gaussian copula is assumed between the friction angle and the cohesion\footnote{This represents the data of \citet{dimatteoLaboratoryShearStrength2013}}. The failure domain is defined by the negative values of $\Mcal_3$ so that the failure function reads $g_3(\xbs) = \Mcal_3(\xbs) - 1$.

\subsubsection{Sheet pile wall $\Mcal_6$}\label{sec_sheet_pile_wall}

\begin{figure}[pos=t!]
\subfloat[\label{fig:Benchmark_WallSketch1}]{\includegraphics[scale=1]{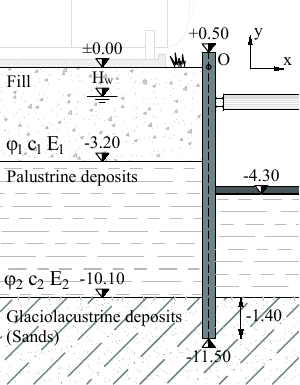}}\hfill
\subfloat[\label{fig:Benchmark_WallSketch2}]{\includegraphics[scale=1]{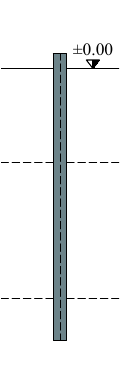}}\hfill
\subfloat[\label{fig:Benchmark_WallSketch3}]{\includegraphics[scale=1]{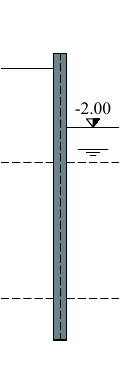}}\hfill
\subfloat[\label{fig:Benchmark_WallSketch4}]{\includegraphics[scale=1]{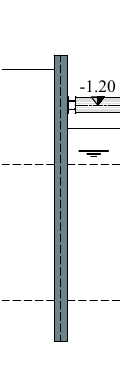}}\hfill
\subfloat[\label{fig:Benchmark_WallSketch5}]{\includegraphics[scale=1]{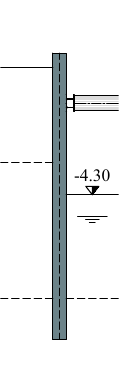}}\hfill
\subfloat[\label{fig:Benchmark_WallSketch6}]{\includegraphics[scale=1]{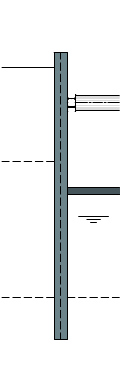}}\hfill
\subfloat[\label{fig:Benchmark_WallSketch7}]{\includegraphics[scale=1]{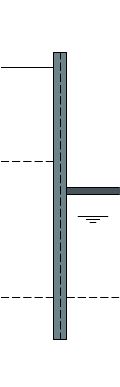}}\hfill
\caption{(a) Sheet pile wall at its final stage with its corresponding stratigraphy. (b to g) Excavation stages with ROR strut and dewatering.}
\label{fig:Benchmark_WallSketch}
\end{figure}

\begin{figure}[pos=b,width=.5\textwidth]
\centering
\captionbox{Vertical displacement field in mm (negative values only) with mean value input parameters\label{fig:Benchmark_WallMean}}[.5\textwidth]{\includegraphics[width=1\linewidth]{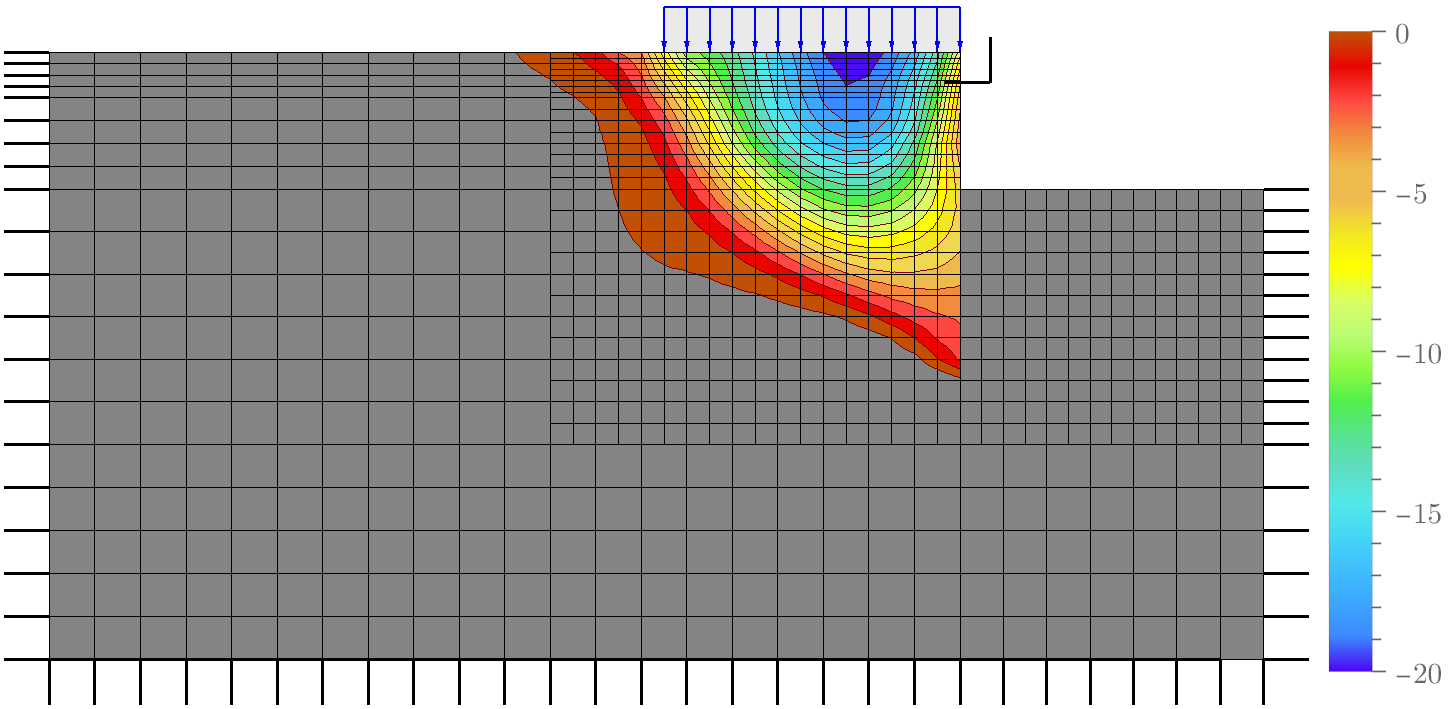}}%
\captionsetup{type=table}\captionbox{Marginal description of the probabilistic model $\Mcal_6$\label{tab:Benchmark_Wall}}[.5\textwidth]{\small
\begin{tabular}{ccccc}
\toprule
\textbf{Variable} & \textbf{Unit} & $f_{\Xbs}$ & $\mu_{\Xbs}$ & $\sigma_{\Xbs}$ \\
\midrule
$\varphi_1$ & $\mathrm{(deg)}$ & $\mathcal{LN}$ & 30.6 & 3.06 \\
$c_1$       & $\mathrm{(kPa)}$ & $\mathcal{LN}$ & 1.0  & 0.20 \\
$E_1$       & $\mathrm{(MPa)}$ & $\mathcal{LN}$ & 11.0 & 2.75 \\
$\varphi_2$ & $\mathrm{(deg)}$ & $\mathcal{LN}$ & 24.0 & 2.40 \\
$c_2$       & $\mathrm{(kPa)}$ & $\mathcal{LN}$ & 7.5  & 1.50 \\
$E_2$       & $\mathrm{(MPa)}$ & $\mathcal{LN}$ & 2.7  & 0.68 \\
$H_w$       & $\mathrm{(m)}$   & $\mathcal{N}$  & 1.0  & 0.10 \\
\bottomrule
\end{tabular}}
\captionsetup{type=figure}
\end{figure}

The next case study is taken from a full-scale project in Yverdon-les-Bains, Switzerland. It involves the construction of a basement in poor soil conditions, including a dewatering excavation. The main challenge is to control deformations to ensure the stability of a sensitive shallowly founded silo at the rear of the excavation. The problem was modelled by 2D finite elements using ZSoil geomechanics software \cite{zsoilrWindowsBasedToolOffering2024}. The non-linear behaviour of the soil is taken into account using the small strain hardening soil model \cite{cudnyRefinementHardeningSoil2020}. A steady-state analysis was considered for the hydromechanical behaviour. The soil stratigraphy and the geometry are both shown in \cref{fig:Benchmark_WallSketch1}. The excavation sequence is carried out as follows: ($t_1$) Sheet pile placement (\cref{fig:Benchmark_WallSketch2}). ($t_2$) First excavation (\cref{fig:Benchmark_WallSketch3}). ($t_3$) Placement of the ROR strut (\cref{fig:Benchmark_WallSketch4}). ($t_4$) Second excavation (\cref{fig:Benchmark_WallSketch5}). ($t_5$) Concreting the bottom slab (\cref{fig:Benchmark_WallSketch6}). ($t_6$) Removal of the ROR strut (\cref{fig:Benchmark_WallSketch7}). Note that the dewatering is performed -0.75m below the excavation bottom. Considering that ZSoil can return the vertical displacement field $u_y(\Xbs,x,y,t)$ in position $(x,y)$ at time $t$, the quantity of interest in the problem is the minimum vertical settlement at the surface of the unexcavated zone : 
\begin{equation}\label{equ:wallQoI}
\Mcal_6(\Xbs,t) = \min_{x<0} u_y(\Xbs,x,0,t) =: u_{y,\text{min}}(\Xbs,t)
\end{equation}

\noindent where $\Xbs \in \Rbb^7$ is composed of seven random variables including the friction angle of the soil $\varphi$, the cohesion $c$, the Young’s modulus $E$ (for both soil layers) as well as the height of the water table $H_w$. The marginal description of the full random vector is described in \cref{tab:Benchmark_Wall}. A negative correlation of -0.7 between the friction angle and the cohesion is assumed and modelled by a Gaussian copula. \cref{fig:Benchmark_WallMean} shows the field of negative vertical displacements calculated from the average of the random vector. The minimal vertical displacement is $-21.1 \mathrm{mm}$.

To ensure the stability of the silo, $u_{y,\text{min}}$ must not exceed 30 mm during the construction, i.e. at step ($t_5$). The failure function is therefore $g_6(\xbs) = 30 - |u_{y,\text{min}}(\xbs,t_5)|$. Because of material non-linearities and contact algorithms, a single model evaluation can take several minutes. Unlike the other models considered in this study, direct computation of the reference probability of failure using crude Monte-Carlo simulation is not viable. Only for estimating the reference reliability index, we used a PCK-based active learning reliability scheme. The algorithm settings were set according to the recommendations of \citet{moustaphaActiveLearningStructural2022a}.

\subsection{Benchmark set-up}\label{sec_Benchmark_procedure}
In order to test our sampling strategy, we apply the systematic workflow illustrated in \cref{fig:Benchmark_FlowChart} to models $\mathcal{M}_2$–$\mathcal{M}_8$. Model $\mathcal{M}_1$ is discussed graphically, and we perform an additional specific analysis for problem $\mathcal{M}_9$. The finite-element model $\Mcal_6$ (wall) further departs from the generic workflow for the RSIE metric: its inputs are correlated, so the reference sensitivity indices are Kucherenko rather than Sobol' indices. As detailed in Section \ref{sec_performance_metrics}, their reference relies on a replicated, adaptively-grown pick-freeze estimation, which would require a number of evaluations incompatible with the minutes-long runtime of a single finite-element call. No reference of controlled accuracy could therefore be established, and $\Mcal_6$ is excluded from the RSIE results; its RRIE is unaffected, as the reference reliability index is obtained by an active-learning scheme. The generic procedure is detailed below for one model $\mathcal{M}_i$. Except for the finite-element model $\mathcal{M}_6$, the problems addressed in this study are computationally inexpensive. After selecting the problem, the workflow proceeds as follows:

\begin{figure}[pos=t!]
\centering
\includegraphics[scale=1]{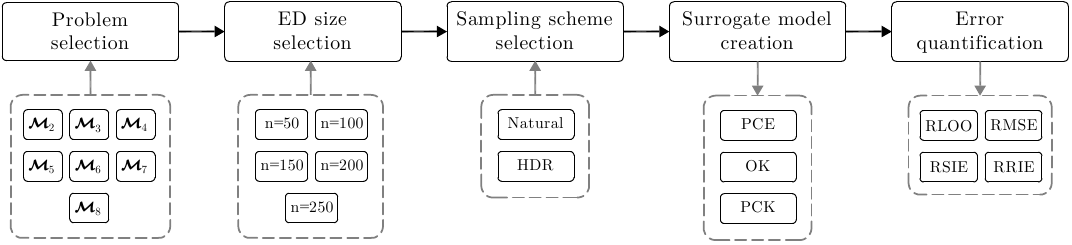}
\caption{Flowchart of our benchmark methodology applied to problems $\Mcal_2$ to $\Mcal_8$.}
\label{fig:Benchmark_FlowChart}
\end{figure}

\begin{enumerate}[label=(\roman*)]
\item We generate a fixed validation set $(\mathcal{X}^{+}, \mathcal{Y}^{+})$ that remains unchanged throughout the analysis. For the finite-element model $\Mcal_6$ we use $m = 10^{4}$ realisations, whereas for the remaining problems we use $m = 10^{6}$.
\item We define the size of the experimental design. In this study we consider five sizes $n=50$, $n=100$, $n= 150$, $n=200$ and $n=250$, chosen according to our expertise.
\item We subsequently generate the experimental design using natural sampling and HDR sampling with probability level $\alpha = 0.01$. In both cases, Monte-Carlo sampling is involved. The sensitivity of the surrogate model to $\alpha$ is discussed in the analysis of model $\mathcal{M}_9$.
\item We construct the three surrogate models described in Section \ref{sec_Surrogate_model_formulation} for both sampling methods.
\item We compute the four error metrics defined in Section \ref{sec_performance_metrics}. As explained above, the RSIE is not evaluated for the finite-element model $\Mcal_6$.
\end{enumerate}

Because the whole process is stochastic by definition, we repeat the entire procedure 100 times to obtain representative error values. The full workflow therefore requires $7 \times (50 + 100 + 150 + 200 + 250) \times 2 \times 3 \times 100 + 6 \times 10^6 + 1 \times 10^4 \approx 9 \times 10^6$ model evaluations and the construction of $7 \times 5 \times 2 \times 3 \times 100 = 21,000$ surrogate models. To achieve a practical runtime, we carried out all simulations on an external server using parallelisation.

\section{Results and discussion}\label{chap_Results}

\subsection{Franke problem $\Mcal_1$}

\begin{figure}[pos=b]
\raggedright \includegraphics[scale = 0.8]{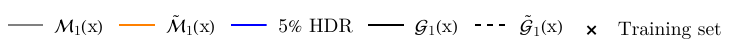}\\
\subfloat[PCE\label{fig:Results_Franke1}\\
]{\includegraphics[scale=1]{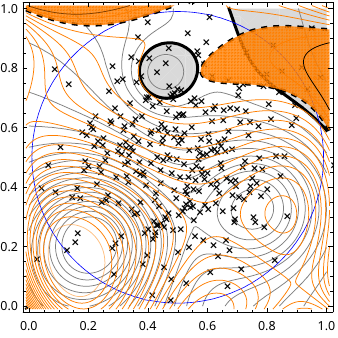}}\hfill
\subfloat[OK\label{fig:Results_Franke3}\\
]{\includegraphics[scale=1]{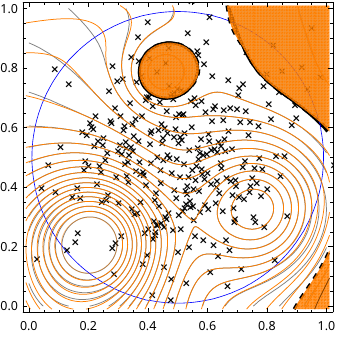}}\hfill
\subfloat[PCK\label{fig:Results_Franke5}\\
]{\includegraphics[scale=1]{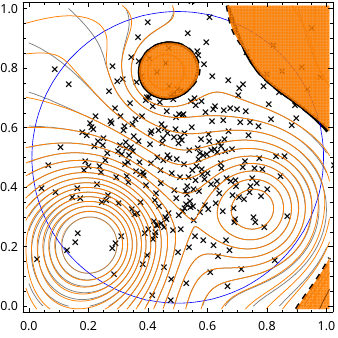}}\\
\subfloat[PCE\label{fig:Results_Franke2}\\
]{\includegraphics[scale=1]{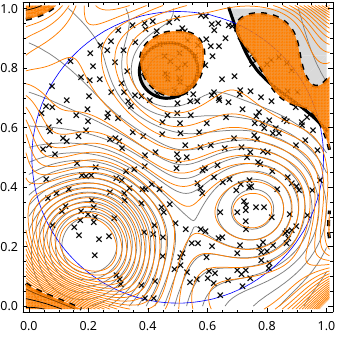}}\hfill
\subfloat[OK\label{fig:Results_Franke4}\\
]{\includegraphics[scale=1]{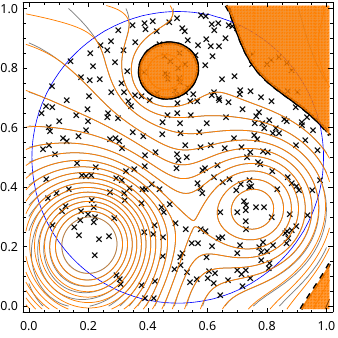}}\hfill
\subfloat[PCK\label{fig:Results_Franke6}\\
]{\includegraphics[scale=1]{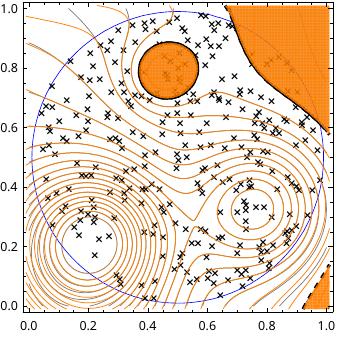}}\vspace{0.3cm}
\centering
\begin{tabular}{llcccc}
\toprule
Method & Distribution & RLOO & RMSE & RSIE & RRIE \\
\midrule
\multirow{2}{*}{PCE} & Natural & $6.68 \times 10^{-2}$ & {\boldmath$1.93 \times 10^{-1}$} & {\boldmath$3.02 \times 10^{-2}$} & $1.23 \times 10^{-1}$ \\
               & HDR & {\boldmath$8.21 \times 10^{-3}$} & $5.08 \times 10^{-1}$ & $2.02 \times 10^{-1}$ & {\boldmath$1.42 \times 10^{-2}$} \\
\midrule
\multirow{2}{*}{Kriging} & Natural & $7.45 \times 10^{-4}$ & {\boldmath$1.43 \times 10^{-3}$} & $5.08 \times 10^{-3}$ & $1.08 \times 10^{-2}$ \\
               & HDR & {\boldmath$4.77 \times 10^{-7}$} & $2.74 \times 10^{-3}$ & {\boldmath$3.58 \times 10^{-3}$} & {\boldmath$8.58 \times 10^{-3}$} \\
\midrule
\multirow{2}{*}{PCK} & Natural & $1.55 \times 10^{-4}$ & {\boldmath$9.36 \times 10^{-4}$} & $6.24 \times 10^{-3}$ & $1.16 \times 10^{-2}$ \\
               & HDR & {\boldmath$4.44 \times 10^{-7}$} & $2.49 \times 10^{-3}$ & {\boldmath$5.78 \times 10^{-4}$} & {\boldmath$1.01 \times 10^{-2}$} \\
\bottomrule
\end{tabular}
\caption{Graphical representations of the approximations of $\Mcal_1$ and their corresponding error magnitudes. (a) to (c) Natural sampling and (d) to (f) HDR sampling.}
\label{fig:Results_Franke}
\end{figure}

First, we compare the two sampling methods using the two-dimensional Franke function $\Mcal_{1}$. Based on an experimental design of size $n = 300$ and an HDR with probability $\alpha = 0.05$, we construct six approximations of the Franke function, as shown in \cref{fig:Results_Franke}. \Cref{fig:Results_Franke1,fig:Results_Franke3,fig:Results_Franke5} correspond to surrogate models trained on a natural design, whereas \cref{fig:Results_Franke2,fig:Results_Franke4,fig:Results_Franke6} correspond to models trained on an HDR design. Although the comparison of surrogate families is not an objective of this study, the graphical results confirm the expected trend: Kriging-based surrogates (OK and PCK) reproduce the contour topology of $\Mcal_1$ with substantially higher fidelity than PCE. These qualitative observations are consistent with the quantitative error values reported in the table in \cref{fig:Results_Franke}. We discuss the four performance metrics separately.

\begin{itemize}

\item[$\varepsilon_{\text{RLOO}}$]
HDR sampling reduces the leave-one-out error for all three surrogates. For PCE, $\varepsilon_{\text{RLOO}}$ decreases by a factor of approximately one order of magnitude ($6.68\times10^{-2} \rightarrow 8.21\times10^{-3}$). For OK, the reduction exceeds three orders of magnitude ($7.45\times10^{-4} \rightarrow 4.77\times10^{-7}$), and for PCK it exceeds two orders of magnitude ($1.55\times10^{-4} \rightarrow 4.44\times10^{-7}$). Thus, HDR markedly improves the predictive capability measured by cross-validation, particularly for Kriging-type surrogates.

\item[$\varepsilon_{\text{RMSE}}$]
In contrast, HDR increases the RMSE for all three surrogates, indicating a systematic loss of global accuracy. The error grows from $1.93\times10^{-1}$ to $5.08\times10^{-1}$ for PCE, from $1.43\times10^{-3}$ to $2.74\times10^{-3}$ for OK, and from $9.36\times10^{-4}$ to $2.49\times10^{-3}$ for PCK. Concentrating the training points within the high-density region thus improves the cross-validation error while degrading the global $L^2$ accuracy assessed over the input distribution.

\item[$\varepsilon_{\text{RSIE}}$]
The relative sensitivity-index error exhibits a surrogate-dependent behaviour. For PCE, HDR is clearly detrimental, with $\varepsilon_{\text{RSIE}}$ increasing by a factor of approximately $7$ ($3.02\times10^{-2} \rightarrow 2.02\times10^{-1}$). Conversely, both Kriging-based surrogates benefit: $\varepsilon_{\text{RSIE}}$ decreases from $5.08\times10^{-3}$ to $3.58\times10^{-3}$ for OK and, more markedly, from $6.24\times10^{-3}$ to $5.78\times10^{-4}$ for PCK, a reduction of roughly one order of magnitude. HDR therefore improves the estimation of the total-order sensitivity indices for the Kriging-type surrogates while penalising the global polynomial approximation provided by PCE.

\item[$\varepsilon_{\text{RRIE}}$]
The relative reliability-index error improves under HDR for all three surrogates, although its magnitude varies considerably. For PCE, HDR yields a clear gain, with $\varepsilon_{\text{RRIE}}$ decreasing by almost one order of magnitude ($1.23\times10^{-1} \rightarrow 1.42\times10^{-2}$). For the Kriging-based surrogates the improvement is more modest: $\varepsilon_{\text{RRIE}}$ decreases from $1.08\times10^{-2}$ to $8.58\times10^{-3}$ for OK and from $1.16\times10^{-2}$ to $1.01\times10^{-2}$ for PCK. These reductions reflect a more accurate representation of the limit-state surface and, consequently, of the failure probability.

\end{itemize}

Overall, this example is a first illustration of the trade-off underlying HDR sampling. By redistributing the experimental design within the high-density region, HDR seems to improve the cross-validation error ($\varepsilon_{\text{RLOO}}$) and the reliability-index error ($\varepsilon_{\text{RRIE}}$), at the expense of the global $L^2$ error ($\varepsilon_{\text{RMSE}}$). The sensitivity error ($\varepsilon_{\text{RSIE}}$) follows the same surrogate-dependent split as the global metrics, improving for the Kriging-type surrogates and degrading for PCE. Note finally that, for PCE, neither design recovers the failure domain accurately: the approximation (orange) departs visibly from the true $\Dcal_f$ (grey) in both \cref{fig:Results_Franke1} and \cref{fig:Results_Franke2}, with HDR yielding a slightly closer match. The Kriging-based surrogates reproduce $\Dcal_f$ closely under both designs, and likewise benefit from HDR.

\subsection{Benchmark problems $\Mcal_2$ to $\Mcal_8$}
\afterpage{\clearpage}

\begin{figure}[p]
\raggedright \includegraphics[scale = 0.8]{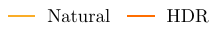}\\
\subfloat[\label{fig:Results_StripFoot1}$\varepsilon_{\text{RLOO}}$]{\includegraphics[scale=1]{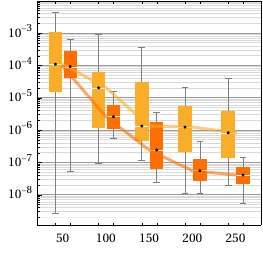}}\hfill
\subfloat[\label{fig:Results_StripFoot2}$\varepsilon_{\text{RMSE}}$]{\includegraphics[scale=1]{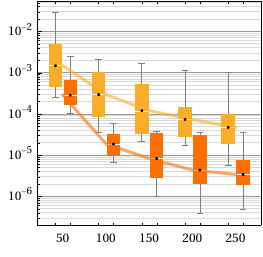}}\hfill
\subfloat[\label{fig:Results_StripFoot3}$\varepsilon_{\text{RSIE}}$]{\includegraphics[scale=1]{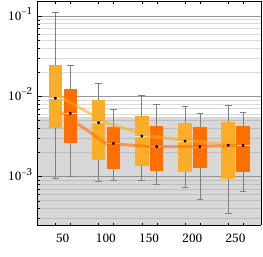}}\hfill
\subfloat[\label{fig:Results_StripFoot4}$\varepsilon_{\text{RRIE}}$]{\includegraphics[scale=1]{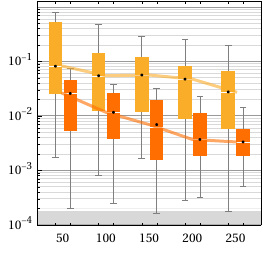}}
\caption{Comparison between natural and HDR sampling over the four error metrics for a PCK-approximation of problem $\Mcal_3$ (strip foundation).}
\label{fig:Results_StripFoot}
\end{figure}

\begin{figure}[p]
\raggedright \includegraphics[scale = 0.8]{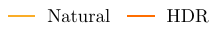}\\
\subfloat[\label{fig:Results_Wall1}$\varepsilon_{\text{RLOO}}$]{\includegraphics[scale=1]{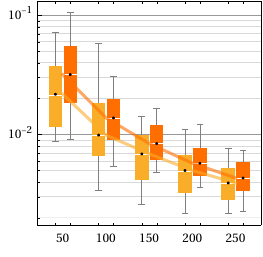}}\hfill
\subfloat[\label{fig:Results_Wall2}$\varepsilon_{\text{RMSE}}$]{\includegraphics[scale=1]{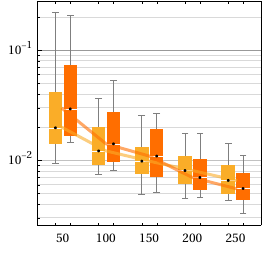}}\hfill
\subfloat[\label{fig:Results_Wall4}$\varepsilon_{\text{RRIE}}$]{\includegraphics[scale=1]{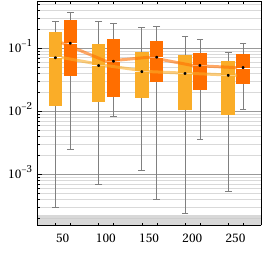}}
\caption{Comparison between natural and HDR sampling over the three error metrics for a PCE-approximation of problem $\Mcal_6$ (retaining wall).}
\label{fig:Results_Wall}
\end{figure}

\begin{figure}[p]
\raggedright \includegraphics[scale = 0.8]{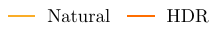}\\
\subfloat[\label{fig:Results_Truss1}$\varepsilon_{\text{RLOO}}$]{\includegraphics[scale=1]{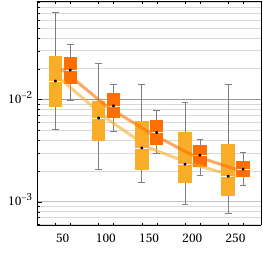}}\hfill
\subfloat[\label{fig:Results_Truss2}$\varepsilon_{\text{RMSE}}$]{\includegraphics[scale=1]{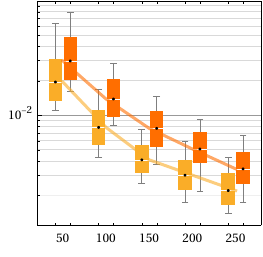}}\hfill
\subfloat[\label{fig:Results_Truss3}$\varepsilon_{\text{RSIE}}$]{\includegraphics[scale=1]{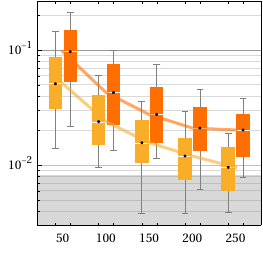}}\hfill
\subfloat[\label{fig:Results_Truss4}$\varepsilon_{\text{RRIE}}$]{\includegraphics[scale=1]{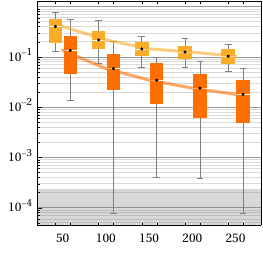}}
\caption{Comparison between natural and HDR sampling over the four error metrics for an ordinary Kriging approximation of problem $\Mcal_8$ (truss structure).}
\label{fig:Results_Truss}
\end{figure}

The Franke example already exhibits the trade-off underlying HDR sampling, but a single problem cannot show how it plays out across applications.

\subsubsection{Individual results ($\Mcal_3$, $\Mcal_6$ and $\Mcal_8$)}

We report three benchmark problems selected to span the observed range of behaviours: the strip foundation $\Mcal_3$ (PCK), for which HDR reduces the error for every task; the retaining wall $\Mcal_6$ (PCE), for which the two designs perform similarly and natural sampling is marginally preferable; and the truss structure $\Mcal_8$ (ordinary Kriging), for which the preferred design depends on the task. \cref{fig:Results_StripFoot,fig:Results_Wall,fig:Results_Truss} show box plots of the four error metrics for the different experimental design sizes. The boxes are defined such that the five-number summary corresponds to the quantiles $0$, $0.1$, $0.5$, $0.9$, and $1$; the curves superimposed on the boxes connect the medians and are centred along the $x$-axis at the midpoint of each pair of boxes. We note that changing the sampling scheme may lead to appreciably different surrogate models, as the LAR algorithm can converge to different multi-index sets and therefore to different selections of basis elements. Whether the two designs are strictly comparable is a legitimate question; our aim, however, is to characterise performance from a user perspective rather than restrict the comparison to a fixed polynomial degree, and we consider this choice appropriate in a practical setting. For the retaining wall, the sensitivity metric is omitted because reliable reference Sobol' indices could not be established for this problem (see \cref{sec_performance_metrics}).

The strip foundation (\cref{fig:Results_StripFoot}) illustrates the regime in which the broader coverage provided by HDR sampling is beneficial across all tasks: HDR yields lower errors than natural sampling for all four metrics and at every experimental design size. The reduction frequently exceeds one order of magnitude; for the global $L^2$ error, the median $\varepsilon_{\mathrm{RMSE}}$ at $n=250$ decreases from $4.7\times 10^{-5}$ to $3.4\times 10^{-6}$ (\cref{fig:Results_StripFoot2}). The reliability metric follows a similar trend: HDR keeps $\varepsilon_{\mathrm{RRIE}}$ below $0.1$ at all design sizes, both in terms of median and maximum values, whereas approximately $40\%$ of the natural-sampling runs exceed this threshold at the smallest design size. The sensitivity metric follows the same trend, with HDR reducing the median $\varepsilon_{\mathrm{RSIE}}$ by a factor of approximately $1.5$ to $1.8$ for $n\leq100$; the two designs become essentially equivalent at $n=150$.

The retaining wall (\cref{fig:Results_Wall}) represents the least favourable case for HDR sampling and provides a useful counterpoint. For $\varepsilon_{\mathrm{RLOO}}$ and $\varepsilon_{\mathrm{RMSE}}$, the two designs perform similarly: natural sampling produces slightly lower errors at smaller design sizes, while HDR catches up and becomes marginally better than natural sampling in terms of $\varepsilon_{\mathrm{RMSE}}$ for $n\geq200$. For $\varepsilon_{\mathrm{RRIE}}$, natural sampling consistently performs better: HDR increases the median error at every design size, by between $17\%$ ($n=100$) and $70\%$ ($n=50$). For both designs, however, the median $\varepsilon_{\mathrm{RRIE}}$ remains in the range of $4\%$--$7\%$ across all design sizes, and essentially no run reaches the noise-only tolerance ($\bar{\varepsilon} \approx 2.2\times10^{-4}$). For this problem, neither one-shot design resolves the limit-state surface sufficiently well for the reliability metric to distinguish between them: both remain well above the reference level, with a relative error on the reliability index too large for direct use in reliability assessment. This is precisely the regime in which active-learning reliability methods are expected to be preferable.

\begin{figure}[pos=b!]
\raggedright \includegraphics[scale = 0.8]{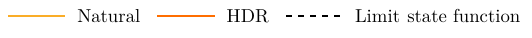}\\
\subfloat[\label{fig:Results_StripFootY}$\Mcal_3$]{\includegraphics[scale=1]{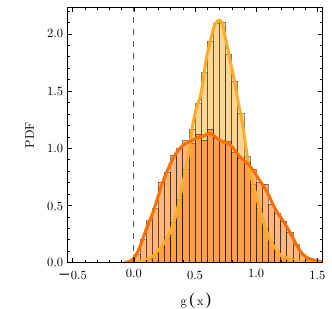}}\hfill
\subfloat[\label{fig:Results_WallY}$\Mcal_6$]{\includegraphics[scale=1]{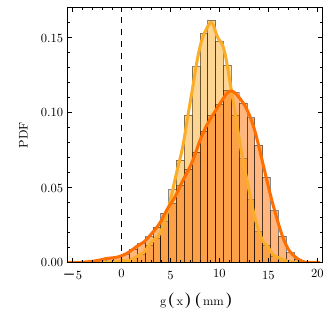}}\hfill
\subfloat[\label{fig:Results_TrussY}$\Mcal_8$]{\includegraphics[scale=1]{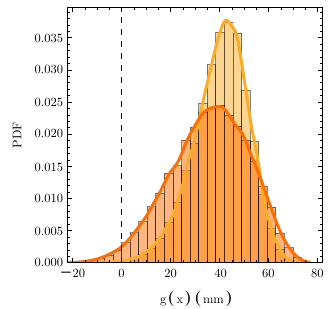}}
\caption{Empirical distributions of the limit-state response $g(\Xbs)$ for problems $\Mcal_3$, $\Mcal_6$, and $\Mcal_8$, evaluated on input samples drawn under natural and HDR sampling ($20{,}000$ samples per design). The vertical line indicates the limit state $g=0$.}
\label{fig:Results_Y}
\end{figure}

The truss structure (\cref{fig:Results_Truss}) is the most instructive case, as the ordering of the two designs depends on the task considered. Natural sampling yields lower errors for $\varepsilon_{\mathrm{RLOO}}$, $\varepsilon_{\mathrm{RMSE}}$, and $\varepsilon_{\mathrm{RSIE}}$ at every design size. At $n=250$, for instance, the median $\varepsilon_{\mathrm{RMSE}}$ is $2.2\times10^{-3}$ for natural sampling, compared with $3.4\times10^{-3}$ for HDR, whereas HDR yields substantially lower errors for $\varepsilon_{\mathrm{RRIE}}$, with the relative reduction increasing from $67\%$ at $n=50$ to $83\%$ at $n=250$. In absolute terms, the median natural-sampling $\varepsilon_{\mathrm{RRIE}}$ remains above $10\%$, whereas HDR falls below this threshold from $n=100$ onwards and reaches approximately $1.8\%$ at $n=250$.

In our view, this ordering reflects a trade-off rather than a contradiction. The three metrics for which natural sampling performs better aggregate the surrogate error under the input density $f_X$, and therefore reward accuracy in the region where $f_X$ is concentrated; a natural design samples this region densely, whereas an HDR design distributes the same budget over a wider support and accepts a coarser approximation near the mode. For the truss structure, this loss remains moderate, with the median $\varepsilon_{\mathrm{RMSE}}$ under HDR exceeding that of natural sampling by a factor of approximately $1.5$ to $1.9$ across design sizes, while the reliability error decreases by more than one order of magnitude. The reliability metric, by contrast, depends on surrogate accuracy near the limit-state surface, which does not necessarily coincide with the high-density region of the input. We therefore interpret these two orderings as two consequences of the same redistribution of sample points, with the relevant criterion determined by the target quantity. The response distributions shown in \cref{fig:Results_Y} allow us to qualitatively examine how this redistribution affects the failure domain. Evaluating the limit-state function on $20{,}000$ samples drawn under each design, the proportion of responses in the failure domain increases from $0.12\%$ under natural sampling to $1.48\%$ under HDR for the truss structure, and from $0.10\%$ to $0.79\%$ for the retaining wall. In both problems, HDR places approximately one order of magnitude more points in the failure domain, which we interpret as direct evidence that its broader support can populate the region governing the reliability estimate. This coverage in $\Dcal_f$ is, however, not sufficient to guarantee a direct reduction in reliability error. The strip foundation suggests a second mechanism leading to the same outcome. In that case, HDR places essentially no samples in the failure domain, compared with $0.03\%$ for natural sampling, yet it still yields lower $\varepsilon_{\mathrm{RRIE}}$ at every design size. As this improvement cannot stem from samples drawn in $\Dcal_f$, we attribute it to the extrapolation capability of the surrogate trained under the HDR design.

\subsubsection{Pooled results}

We now pool the results across all problems and design sizes to compare the two designs at the scale of the whole benchmark. Each pair (problem, design size) is treated as a \textit{heat}. In every heat the two sampling schemes are represented by their $100$ replications, and the $200$ resulting runs are ranked jointly according to the error of interest. The joint ranking is grouped into ordered categories corresponding to the empirical quantiles of the pooled errors: a best category, then the $5\%$, $10\%$, $25\%$ and $50\%$ quantiles, and the remainder. For $\varepsilon_{\mathrm{RLOO}}$ and $\varepsilon_{\mathrm{RMSE}}$ the ranking follows the error value directly, so that the best category contains the single most accurate run of the heat. The two threshold-based metrics, $\varepsilon_{\mathrm{RRIE}}$ and $\varepsilon_{\mathrm{RSIE}}$, receive a specific treatment: every run whose error falls at or below the noise-only tolerance is regarded as having reached reference-level accuracy and is placed in the best category and might share it ex~aequo, while the remaining runs are ranked across the usual quantile categories. This construction also quantifies, over the entire study, the proportion of one-shot surrogates that attain reference-level accuracy. Considering $200$ runs over seven problems and five design sizes, $3500$ ranks are assigned to each combination of surrogate and design. As before, problem $\Mcal_6$ is excluded from $\varepsilon_{\mathrm{RSIE}}$, leaving a total of $3000$ ranks.

\begin{figure}[pos=t!]
\raggedright \includegraphics[scale = 0.8]{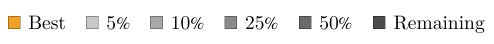}\\
\subfloat[\label{fig:Results_Benchmark1}$\varepsilon_{\text{RLOO}}$]{\includegraphics[scale=1]{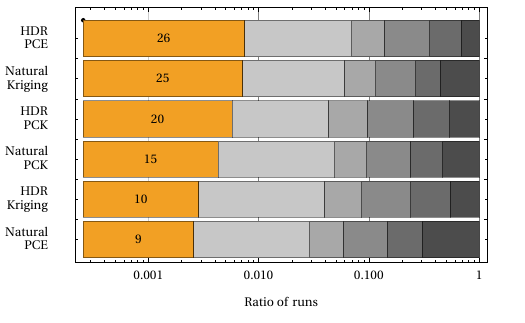}}\hfill
\subfloat[\label{fig:Results_Benchmark2}$\varepsilon_{\text{RMSE}}$]{\includegraphics[scale=1]{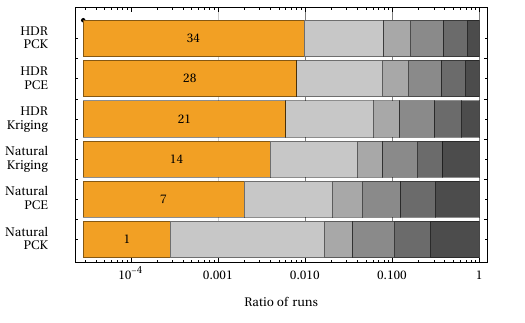}}\\
\subfloat[\label{fig:Results_Benchmark3}$\varepsilon_{\text{RSIE}}$]{\includegraphics[scale=1]{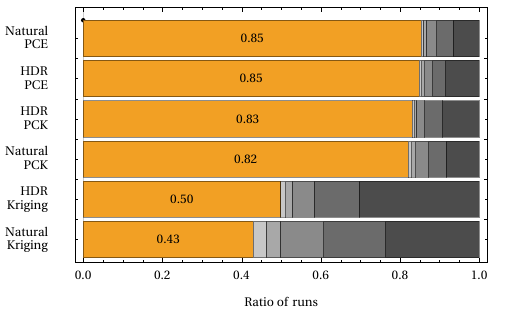}}\hfill
\subfloat[\label{fig:Results_Benchmark4}$\varepsilon_{\text{RRIE}}$]{\includegraphics[scale=1]{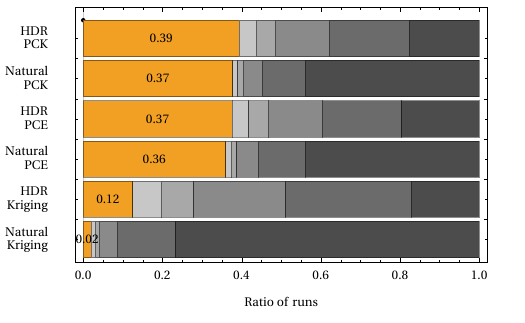}}
\caption{Distribution of the runs across the ranking categories for the four error metrics. For each metric, the six surrogate--design combinations are sorted by the share of runs in the best category.}
\label{fig:Results_Benchmark}
\end{figure}

\cref{fig:Results_Benchmark} reports, for the four metrics, the distribution of the six surrogate--design combinations across the ranking categories, each metric being sorted by the share of runs in the best category. For the predictive metrics, HDR sampling shifts the distribution of outcomes towards the better categories for PCE and PCK (\cref{fig:Results_Benchmark1,fig:Results_Benchmark2}). The most accurate run of a heat is obtained with HDR sampling in $46$ of the $70$ PCE--PCK heats for $\varepsilon_{\mathrm{RLOO}}$ and in $62$ for $\varepsilon_{\mathrm{RMSE}}$, and the lighter categories are correspondingly more populated under HDR. The shift is also visible at the other end of the ranking: for PCE and PCK, natural sampling accumulates most of its runs in the worst category, holding about $62\%$ of them for $\varepsilon_{\mathrm{RLOO}}$ and $71\%$ for $\varepsilon_{\mathrm{RMSE}}$, against $38\%$ and $29\%$ under HDR. For ordinary Kriging the picture is more nuanced: natural sampling provides the most accurate run more often for $\varepsilon_{\mathrm{RLOO}}$ ($25$ of $35$ heats), whereas HDR does so for $\varepsilon_{\mathrm{RMSE}}$ ($21$ of $35$). The worst category is more balanced for this surrogate, the two designs being close for $\varepsilon_{\mathrm{RLOO}}$ while HDR remains less represented for $\varepsilon_{\mathrm{RMSE}}$. For the sensitivity metric (\cref{fig:Results_Benchmark3}), the proportion of runs reaching reference-level accuracy is essentially identical for the two designs in the case of PCE and PCK (about $0.85$ and $0.82$ to $0.83$, respectively). Both designs sit close to this ceiling, so $\varepsilon_{\mathrm{RSIE}}$ does not separate them for these surrogates, and the worst category is marginal under either design (below $10\%$ of the runs). For ordinary Kriging the proportion reaching reference level is somewhat higher under HDR ($0.50$ against $0.43$), but so is the worst category ($30\%$ against $24\%$), the gain at the top being partly offset by a slightly heavier lower tail. At the level of individual problems the ordering can go either way, the truss structure being a case where natural sampling is favoured. We therefore interpret the pooled sensitivity results as indicating that HDR retains the accuracy of the total-order indices on average, with a modest improvement for ordinary Kriging, rather than as a systematic advantage of one design over the other. The reliability error follows a comparable pattern (\cref{fig:Results_Benchmark4}). For PCE and PCK the proportion of runs reaching reference-level reliability is comparable for the two designs, marginally higher under HDR ($0.37$ against $0.36$ for PCE, $0.39$ against $0.38$ for PCK). The two designs separate more clearly at the bottom of the ranking, where natural sampling accumulates close to $44\%$ of its runs in the worst category for both surrogates, against about $18\%$ to $20\%$ under HDR. For ordinary Kriging the difference is more pronounced, HDR reaching reference-level reliability in $0.12$ of the runs against $0.02$ for natural sampling, while $73\%$ of the natural-sampling runs fall in the worst category against $17\%$ under HDR, in line with the truss structure example.

Taken together, the pooled results indicate that HDR sampling yields, on average, surrogates comparable to, or more accurate than, those obtained with natural sampling on most surrogate-task combinations. The clearest gains are in predictive accuracy for PCE and PCK and in reference-level reliability for ordinary Kriging, while sensitivity accuracy is retained throughout. Where natural sampling leads, namely the upper tail of leave-one-out accuracy for ordinary Kriging and individual sensitivity cases such as the truss structure, the absolute differences remain small. We see these results as supporting the use of HDR as a one-shot design strategy whose benefits extend across a range of UQ tasks. Our conclusions are necessarily tied to the set of problems considered. The seven benchmarks were chosen as documented, practice-oriented engineering models whose behaviour is amenable to interpretation, which is what allowed the reading proposed above. They span only a limited portion of the relevant feature space, however, in terms of input dimensionality, magnitude of the failure probability and limit-state geometry, and different conclusions could well be reached on other classes of problems. A natural extension of this work would be to reproduce the benchmark on a larger and more diverse set, such as the reliability problems collected in \citep{rozsasComputationalChallengesReliability2018}, to assess how broadly the present trends hold.

\subsection{$d$-dimensional problem $\Mcal_9$ (discussing $\alpha$)}

\begin{figure}[pos=b!]
\raggedright \includegraphics[scale = 0.8]{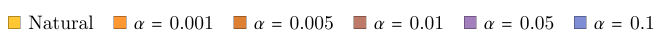}\\
\subfloat[\label{fig:Results_Toy1.pdf}$\varepsilon_{\text{RLOO}}$]{\includegraphics[scale=1]{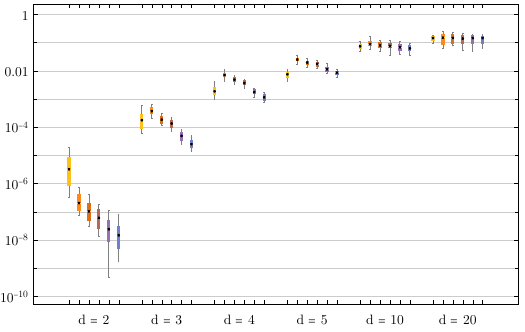}}\hfill
\subfloat[\label{fig:Results_Toy2.pdf}$\varepsilon_{\text{RMSE}}$]{\includegraphics[scale=1]{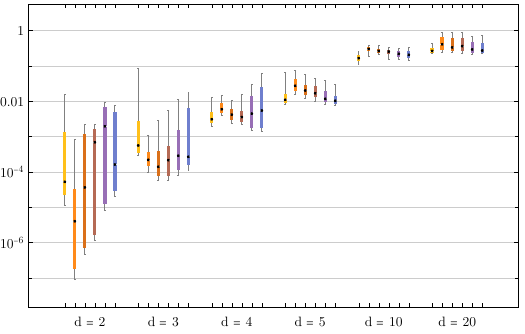}}\\
\subfloat[\label{fig:Results_Toy3.pdf}$\varepsilon_{\text{RRIE}}$ with $P_f = 10^{-2}$]{\includegraphics[scale=1]{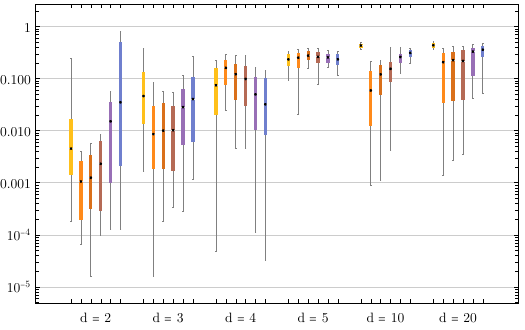}}\hfill
\subfloat[\label{fig:Results_Toy4.pdf}$\varepsilon_{\text{RRIE}}$  with $P_f = 10^{-6}$]{\includegraphics[scale=1]{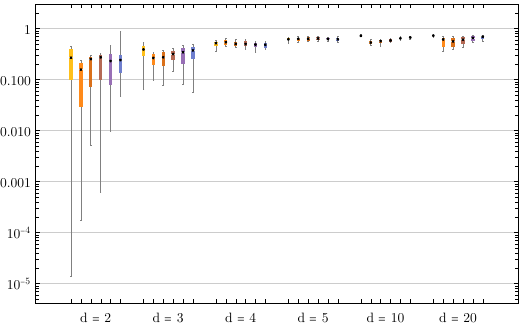}}
\caption{Comparison between natural and HDR sampling with several $\alpha$-values over the three error metrics for a PCK-approximation of problem $\Mcal_9$.}
\label{fig:Results_Toy}
\end{figure}

We have thus far tested our approach on various problems with dimensionalities ranging from $d = 3$ to $d = 10$. The analyses were performed using a fixed probability level $\alpha = 10^{-2}$, defined a priori for the construction of the HDR. We now investigate the influence of this parameter on a PCK approximation of the analytical problem $\Mcal_9$, using $n = 200$ training points for dimensions ranging from $d = 2$ to $d = 20$. We recall that our current implementation, which relies on an acceptance–rejection scheme, is affected by the curse of dimensionality; consequently, dimensions higher than $d>20$ were not considered.

\cref{fig:Results_Toy} shows box plots obtained from 100 repetitions for each sampling scheme, using $n = 200$ samples drawn from either natural- or HDR sampling, for probability levels $0.001$, $0.005$, $0.01$, $0.05$, and $0.1$. The five-number summary of these box plots is defined as in the previous section. As for the Franke function, we discuss the results by separating the metrics:

\begin{itemize}
\item[$\varepsilon_{\text{RLOO}}$]
A systematic decrease in the RLOO error is observed as the level $\alpha$ increases, although this effect tends to be less pronounced with increasing $d$. This trend results from the contraction of the training domain $\Dcal_\ell$: For a fixed sample size $n$, denser sampling improves local coverage. On a smaller domain, the underlying target function becomes locally smoother and the polynomial component of the PCK is numerically more stable, potentially leading to a lower RLOO. As $\alpha \to 1.0$, $\Dcal_\ell$ converges to a single point; the RLOO then essentially measures in-sample reconstruction capability over a highly densified region, producing an optimistic metric with no guarantee of generalisation to the full domain of the target law $f_X$. In practice, the RLOO should be interpreted with caution when HDR-based sampling is used, particularly for large $\alpha$ values, as this may yield overly optimistic conclusions and underestimate the prediction error outside $\Dcal_\ell$ although our method is obviously designed to be applied with small values of $\alpha$.

\item[$\varepsilon_{\text{RMSE}}$]
The RMSE metric clearly demonstrates the degradation of HDR sampling performance compared to natural sampling as dimensionality increases. Although the context differs, these results are consistent with those reported by \citet{luthenSparsePolynomialChaos2021a}, who showed that the performance of a PCE trained with uniform sampling in a hypersphere of radius depending on the target polynomial degree \citep{hamptonCoherenceMotivatedSampling2015} deteriorates with increasing dimension. In our case, natural sampling outperforms HDR for any tested value of $\alpha$ once $d > 3$. We stress again that the generation of the validation set $\Xcal^{+}$ introduces a bias into this metric that inherently penalises HDR, since the discrepancy between $\Mcal$ and $\widetilde{\Mcal}$ is more frequently evaluated near the mode of the distribution, which, in this case, coincides with the mean of the random vector.

\item[$\varepsilon_{\text{RRIE}}$]
HDR sampling outperforms natural sampling for most of the tested values of $\alpha$. In low dimensions, the probability level $\alpha$ appears to have a stronger influence on the error than in higher dimensions. In \cref{fig:Results_Toy3.pdf}, a pronounced error reduction is observed in high dimensions for small values of $\alpha$. A similar trend is seen in \cref{fig:Results_Toy4.pdf}, although neither method yielded a sufficiently accurate approximation of the failure probability, as indicated by the large magnitude of $\varepsilon_{\text{RRIE}}$ across all tested dimensions. This highlights the relevance of using active learning methods, such as those proposed by \cite{moustaphaActiveLearningStructural2022a}, to better approximate $\Dcal_f$ in case of small target probabilities. Because of their fundamentally different nature, we deliberately chose not to compare our method with sequential approaches which is also supported by our strong expectation that sequential methods would outperform the proposed one-shot sampling. Nevertheless, as an alternative and for future work, we propose to combine our approach with an active learning scheme to assess the convergence rate of an adaptive workflow in which the initial experimental design is drawn from natural sampling or HDR.
\end{itemize}

As a final remark, we emphasise that we do not address the optimal choice of $\alpha$, as it may strongly depend on the specific problem or on the nature of the random vector under consideration. As a guideline, we suggest the following reasoning for practitioners: when aiming to obtain, for instance, $n = 100$ samples of a random vector, the probability of obtaining a higher pointwise coverage using HDR sampling with a level $\alpha = 0.01$ compared to natural sampling is, by definition, only 1\%. Thus, the gain in information about the underlying model is potentially limited compared to the gain achieved through uniform coverage over $\Dcal_\ell$. Since surrogate models are typically trained on experimental designs with low cardinality, we are inclined to favour the use of HDR sampling, particularly when its final application involves the surrogate in a predictor-mode.

\section{Conclusion}\label{chap_Conclusion}

This contribution investigated the use of an alternative sampling scheme for training polynomial chaos expansion, ordinary Kriging and polynomial chaos Kriging surrogate models. Based on a heuristic and geometric criterion, we proposed to sample uniformly within the compact support defined by the highest-density region of the input random vector. The proposed scheme was compared to natural sampling, i.e. sampling according to the input distribution, through a benchmark study involving nine problems, including analytical functions, structural engineering applications, and geomechanics problems incorporating finite element simulations. The comparison relied on four task-oriented error metrics: the relative leave-one-out and relative mean square errors, which measure predictive accuracy; the relative sensitivity index error, which quantifies the accuracy of the total-order Sobol' indices; and the relative reliability index error, which measures the distance between the approximated and reference probability of failure. This study led to the following key observations:

\begin{enumerate}[label=(\roman*)]
\item The proposed sampling method is based solely on the original input distribution in order to obtain an HDR. It places no condition on the form of that distribution and therefore applies regardless of the input law. It offers practical flexibility when it comes to actually sampling, as instead of Monte Carlo uniform sampling, other methods might be used such as LHS, Sobol or even sparse grids. The scheme is moreover parallelisable, and can be integrated into a wide range of existing surrogate-modelling frameworks.

\item Within the benchmark considered, HDR sampling yielded surrogates that were, on most surrogate-task combinations, equal to or more accurate than those trained on natural sampling across the four metrics. The gains were clearest in predictive accuracy for PCE and PCK and in reliability for ordinary Kriging. On the combinations where natural sampling led, the absolute differences remained practically acceptable.

\item Because it is produced once and is not tied to a single quantity of interest, an HDR design yields a surrogate that can be reused across several downstream tasks. In its present form, however, the method is restricted to low-dimensional problems, as the acceptance-rejection construction of the design becomes increasingly inefficient as the dimension grows.
\end{enumerate}

For future studies, we strongly recommend an extensive validation of our approach on problems of different nature and application fields. In particular, we encourage work addressing the following two aspects:

\begin{enumerate}[label=(\roman*)]
\item The uniform sampling within the highest-density region currently relies on an acceptance-rejection scheme, whose efficiency degrades in high dimension. Proposing a direct construction of the design that avoids this step would lift the present dimensionality limitation of the method.

\item Applying the scheme to a wider and more diverse set of problems would test whether the trade-off observed here holds across other applications. This includes extending the comparison to further task-oriented metrics, for instance those arising in optimisation or Bayesian inference, so as to probe the behaviour of HDR sampling beyond the three uses considered in this study.
\end{enumerate}

\section*{CRediT authorship contribution statement}
\textbf{Jocelyn Minini}: Conceptualization, Methodology, Software, Validation, Writing - Original Draft, Visualization. \textbf{Micha Wasem}: Conceptualization, Formal analysis, Writing - Review \& Editing, Supervision, Funding acquisition

\section*{Declaration of competing interest}
The authors declare that they have no known competing financial interests or personal relationships that could have appeared to influence the work reported in this paper.

\section*{Data availability}
Data will be made available upon request.

\section*{Acknowledgements}
We warmly thank our colleague Corinne Hager J\"orin for her valuable inputs and feedback regarding the probabilistic and statistical aspects of this study.

\section*{Funding source}
This study has been financed by the ``Ingénierie et Architecture'' domain of HES-SO, University of Applied Sciences Western Switzerland, which is acknowledged.

\newpage
\appendix
\section{Highest density regions of multivariate normal distributions}\label{sec_appendix_gaussian}

Consider a $d$-dimensional normal distribution $\mathcal{N}(\boldsymbol{\mu},\boldsymbol{\Cbf})$ with covariance matrix $\Cbf\in\Rbb^{d\times d}$ which is symmetric and assumed to be positive definite. Its probability density function reads 
\[
f(\xbs) = \frac{1}{(2\pi)^{d/2}\sqrt{\det(\Cbf)}} \exp\left(-\frac{1}{2}(\xbs-\boldsymbol{\mu})\transpose\Cbf^{-1}(\xbs-\boldsymbol{\mu})\right).
\]
The highest density region of level $\alpha$ is given by \(\Dcal_{\ell}\) defined as:
\[
\Dcal_{\ell} = \{\xbs \in \mathbb{R}^{d} : f(\xbs) \geq \ell \},
\]
where $\ell$ is chosen such that
\[
\idotsint_{\Dcal_{\ell}} f(\xbs)\,\mathrm d\xbs = 1 - \alpha.
\]
\noindent We have the following
\begin{prop}
If $S^{d-1}_r$ denotes the $(d-1)$-dimensional sphere of radius $r$ and $\chi^2_{d,1-\alpha}$ is the $(1-\alpha)$-quantile of a $\chi^2_d$ distribution, then for a given $\alpha$, the corresponding value of $\ell$ is given by
\[
\ell = \frac{1}{(2\pi)^{d/2}\sqrt{\det(\Cbf)}} \exp\left(-\frac{\chi^2_{d,1-\alpha}}{2}\right).
\]
and the boundary $\partial\Dcal_{\ell}$ is given by
\[
\mathbf{L}\left(S^{d-1}_{r(\alpha)}\right)+\boldsymbol{\mu},
\]
where $\Lbf$ denotes the Cholesky factor in the Cholesky decomposition $\Cbf = \Lbf \Lbf\transpose$ and $r(\alpha) = \sqrt{\chi^2_{d,1-\alpha}}$. 
\end{prop}

\begin{proof}
Let $
(\xbs-\boldsymbol{\mu})^\mathrm{T}\Cbf^{-1}(\xbs-\boldsymbol{\mu}) = c.$
If we fix a level \(\ell\), we obtain
\[
f(\xbs) \geq \ell \iff \frac{1}{(2\pi)^{d/2}\sqrt{\det(\Cbf)}} \exp\left(-\frac{c}{2}\right) \geq \ell.
\]
From this inequality, we can solve for \(c\) and obtain
\[
c \leq  -2\ln\Bigl(\ell (2\pi)^{d/2}\sqrt{\det(\Cbf)}\Bigr).
\]

\noindent
Using the Cholesky decomposition, we write $\Cbf= \mathbf L\mathbf L\transpose$, where $\mathbf L$ is lower triangular with positive entries on its diagonal. Then we can standardise the normal variable $\Xbs \sim \mathcal N(\boldsymbol{\mu},\Cbf)$ by setting
$
\Ybs = \mathbf L^{-1}(\Xbs-\boldsymbol{\mu}).
$
Therefore
\[
(\Xbs-\boldsymbol{\mu})\transpose\Cbf^{-1}(\Xbs-\boldsymbol{\mu}) = \Ybs\transpose\Ybs \sim \chi^2_d
\]
follows a chi-squared distribution with $d$ degrees of freedom.
\noindent
The condition
\[
\idotsint_{\Dcal_{\ell}} f(\xbs)\,\mathrm d\xbs = 1 - \alpha,
\]
might be recast as $
\Pbb(f(\Xbs) \geq \ell) = 1 - \alpha.$
\noindent
Since \(f(\Xbs) \geq \ell\) corresponds to $
(\Xbs-\boldsymbol{\mu})^{T}\Cbf^{-1}(\Xbs-\boldsymbol{\mu}) \leq c,$
we have:
\[
\Pbb((\Xbs-\boldsymbol{\mu})\transpose\Cbf^{-1}(\Xbs-\boldsymbol{\mu}) \leq c) = 1 - \alpha.
\]
\noindent
Using the \(\chi^2_d\) distribution this reads
\(
\Pbb(\Ybs\transpose\Ybs\leq c) = 1 - \alpha.
\)
Hence $
c = \chi^2_{d,1-\alpha},$
the \((1-\alpha)\)-quantile of the \(\chi^2_d\) distribution.
Putting everything together, we obtain
\[
\ell = \frac{1}{(2\pi)^{d/2}\sqrt{\det(\Cbf)}} \exp\left(-\frac{\chi^2_{d,1-\alpha}}{2}\right)
\]
and the level set $\partial\mathcal D_\ell$ can be explicity described as
$
\mathbf{L}\left(S^{d-1}_{r(\alpha)}\right)+\boldsymbol{\mu},
$
where $S^{d-1}_r$ denotes the $d-1$-dimensional sphere of radius $r$ and $r(\alpha) = \sqrt{\chi^2_{d,1-\alpha}}$.
\end{proof}

\newpage
\section{Custom $d$-dimensional function}\label{sec_appendix_d_dimensional}

Let $h:\Rbb\to\Rbb$ be an injective function which will wlog be assumed to be strictly increasing and let
$g(\xbs) = h\left(\sum_{i=1}^d x_i\right) - C(d)$ where $C(d)$ is a dimensional constant contained in the image of $h$.

\begin{prop} The integral of a standard Gaussian density $f$ on $\{g<0\}$ equals $0<P_f<1$ if and only if
\[C(d) = h\left(\sqrt d\cdot \Phi^{-1}(P_f)\right).\]
\end{prop}
\begin{proof}
The condition $g<0$ is equivalent to
\[
h\left(\sum_{i=1}^d x_i\right)<C(d)
\Longleftrightarrow
\sum_{i=1}^d x_i < h^{-1}(C(d)),
\]
as $C(d)$ is assumed to be in the image of $h$.
The integral

\[
P_f=\idotsint_{\left\{g<0\right\}}f_{\Xbs}(\xbs)\,\mathrm d\xbs
\]
can be interpreted as $\Pbb\left(\sum_{i=1}^d X_i < h^{-1}(C(d))\right)$, where each $X_i$ follows a standard normal distribution. Since $Y=\sum_{i=1}^dX_i$ follows a normal distribution with mean $0$ and variance $\sigma^2=d$, we obtain
\[\begin{aligned}
P_f=\idotsint_{\left\{g<0\right\}}f_{\Xbs}(\xbs)\,\mathrm d\xbs = \Pbb(Y<h^{-1}(C(d))) = \int_{-\infty}^{h^{-1}(C(d))} f_Y(x)\,\mathrm dx,
\end{aligned}\]
where $f_Y$ is the density of $Y$ and hence
\[
P_f=\Phi\left(\frac{h^{-1}(C(d))}{\sqrt d}\right).
\]
Solving for $C(d)$ yields
$
C(d) = h\left(\sqrt d\cdot \Phi^{-1}(P_f)\right).$
\end{proof}

\section{Reference sensitivity indices}\label{sec_appendix_refsensitivity}


\begin{table}[pos=h]
\centering
\footnotesize
\setlength{\tabcolsep}{4pt}
\caption{Reference total-effect sensitivity indices $S_{T,i}$ of inputs $X_i$ for problems $\Mcal_1$--$\Mcal_8$.}
\label{tab:refSensitivity}
\resizebox{\textwidth}{!}{%
\begin{tabular}{ll*{10}{c}}
\toprule
Acronym & Name & $S_{T,1}$ & $S_{T,2}$ & $S_{T,3}$ & $S_{T,4}$ & $S_{T,5}$ & $S_{T,6}$ & $S_{T,7}$ & $S_{T,8}$ & $S_{T,9}$ & $S_{T,10}$ \\
\midrule
$\Mcal_1$ & Franke function & 0.311 & 0.790 &  &  &  &  &  &  &  &  \\
$\Mcal_2$ & Short column & 0.271 & 0.118 & 0.270 &  &  &  &  &  &  &  \\
$\Mcal_3$ & Strip foundation & 0.208 & 0.199 & 0.108 & 0.641 &  &  &  &  &  &  \\
$\Mcal_4$ & Bracket structure & 0.582 & 0.000 & 0.349 & 0.000 & 0.070 &  &  &  &  &  \\
$\Mcal_5$ & Infinite slope & 0.000 & 0.702 & 0.240 & 0.064 & 0.001 & 0.004 &  &  &  &  \\
$\Mcal_6$ & Sheet pile wall & 0.078 & 0.001 & 0.101 & 0.177 & 0.115 & 0.284 & 0.292 &  &  &  \\
$\Mcal_7$ & Steel column & 0.592 & 0.016 & 0.054 & 0.054 & 0.002 & 0.210 & 0.000 & 0.076 & 0.000 &  \\
$\Mcal_8$ & Truss structure & 0.371 & 0.013 & 0.371 & 0.013 & 0.005 & 0.037 & 0.078 & 0.078 & 0.038 & 0.005 \\
\bottomrule
\end{tabular}%
}
\end{table}

\newpage
\bibliographystyle{cas-model2-names}
\bibliography{main.bib}

\end{document}